\documentclass{aa}

\usepackage{graphicx}
\usepackage{txfonts}
\usepackage{natbib,twoopt}
\bibpunct{(}{)}{;}{a}{}{,} %% natbib format for A&A and ApJ
\makeatletter

\begin{document}

   \title{Lower limit for differential rotation in members of young loose stellar associations}

   \author{E. Distefano
          \inst{1}
          \and
          A. C. Lanzafame
          \inst{2}
          \and
          A. F. Lanza\inst{1}
          \and S. Messina\inst{1}
          \and F. Spada\inst{3}
         }

   \institute{ INAF - Osservatorio Astrofisico di Catania\\
              Via S. Sofia, 78, 95123, Catania, Italy\\
              \email{elisa.distefano@oact.inaf.it}
            \and University of Catania, Astrophysics Section, Dept. of Physics and Astronomy\\
          Via S. Sofia, 78, 95123, Catania, Italy  
          \and Leibniz-Institut f\"ur Astrophysik Potsdam (AIP)\\
           An der Sternwarte 16, D-14482, Potsdam, Germany \\         
             }

   \date{Received ; accepted }

% \abstract{}{}{}{}{} 
% 5 {} token are mandatory
 
  \abstract
   { Surface differential rotation (SDR) plays a key role in dynamo models and determines a lower limit on the accuracy of stellar rotation period measurements.  SDR estimates are therefore essential for constraining  theoretical models and inferring realistic rotation period uncertainties. }
   {We  measure a lower limit to SDR in a sample of solar-like stars belonging to young loose stellar associations with the aim of investigating how SDR depends on global stellar parameters in the age range ($4-95~\rm Myr$).}
   { The rotation period of a solar-like star can be recovered by analyzing the flux modulation caused by dark spots and stellar rotation. The SDR and the latitude migration of dark-spots induce  a modulation of the detected rotation period.  We employ long-term photometry to measure the amplitude of such a modulation and to compute the quantity $\Delta\Omega_{\rm phot} =\frac{2\pi}{P_{\rm min}} -\frac{2\pi}{P_{\rm max}}$ that is a lower limit to SDR.
 }
   {We find that $\Delta\Omega_{\rm phot}$ increases with the stellar effective temperature and with the global convective turn-over time-scale $\tau_{c}$, that is the characteristic time for the rise of a convective element through the stellar convection zone. We find that  $\Delta\Omega_{\rm phot}$ is proportional to $T_{\rm eff}^{2.18\pm 0.65}$ in stars recently settled on the ZAMS. This power law is less steep than those found by previous authors, but closest to recent theoretical models. 
   We investigate how $\Delta\Omega_{\rm phot}$ changes in time in   $\sim 1~\rm M_{\sun}$ star. We find that  $\Delta\Omega_{\rm phot}$  steeply increases between 4 and 30 Myr and that  itis almost constant between 30 and 95 Myr.
   We find also that the relative shear increases with the Rossby number $Ro$. Although our results are qualitatively in agreement with  hydrodynamical mean-field models, our measurements are systematically higher than the values predicted by these models. The discrepancy between  $\Delta\Omega_{\rm phot}$ measurements  and  theoretical models is particularly large in stars with periods between 0.7 and 2 d. Such a discrepancy, together with the anomalous SDR measured by other authors for HD 171488 (rotating in 1.31 d), suggests that the rotation period could influence SDR more than predicted by the models.   }
{}


   \maketitle
%
%________________________________________________________________
\section{Introduction}
 Surface differential rotation (here after SDR)  has been detected and measured in a wide sample of solar-like stars   \citep[e.g.][]{2005MNRAS.357L...1B,2007AN....328.1030C,2013A&A...560A...4R}.
These measurements  are  essential to constrain theoretical models that try to investigate the connections between   rotation,  convection and  topology of  stellar magnetic fields.
The amplitude of SDR influences the kind of dynamo operating inside the star.
Indeed, differential rotation is able to generate a strong toroidal magnetic field out a weak poloidal field  and, therefore, is a key ingredient of $\alpha\Omega$ dynamos \citep[e.g.][]{1955ApJ...122..293P,1991GApFD..61..179B}.
On the other hand, stars characterized by a low SDR degree and rotating as solid bodies can instead sustain only  an $\alpha^2$-type dynamo \citep[e.g.][]{1995GApFD..80..229M,2011AN....332..933K}. 
\par The amplitude of SDR  is usually  measured through the surface rotational shear:
\begin{equation}
\Delta \Omega = \Omega_0 - \Omega_{\rm pole}
\end{equation}
where $\Omega_0$ and $\Omega_{\rm pole}$ are the surface angular velocities at equator and at poles, respectively.  If $\Delta\Omega >0$ then the star has a solar-like SDR, i.e. the equator rotates faster than poles. If $\Delta\Omega <0$ the star exhibits an anti-solar SDR.
\par In recent years, several authors investigated how SDR depends on stellar effective temperature and rotation period.
\citet{2005MNRAS.357L...1B} report   $\Delta\Omega$ measurements obtained by means of the Doppler imaging technique for a sample of ten young late-type stars.    
They  find that $\Delta\Omega$ is strongly dependent on the stellar temperature ($\Delta\Omega \propto T_{\rm eff}^{8.93\pm0.31}$) but poorly correlated to $\Omega_0$ ($\Delta\Omega \propto \Omega_0^{0.15\pm0.1}$). A strong correlation between $\Delta\Omega$ and $T_{\rm eff}$ is also found by \citet{2007AN....328.1030C} who enlarges the sample of stars studied by \citet{2005MNRAS.357L...1B} and finds the relationship $\Delta\Omega  \propto T_{\rm eff}^{8.6}$.
\par The results found by these  works are in quantitative disagreement  with  theoretical models developed by  \citet{1999A&A...344..911K}, \citet{2011AN....332..933K} and by \citet{2011A&A...530A..48K}. Such theoretical works predict that SDR increases with the stellar temperature and that it is almost independent on the stellar rotation period. However,  the power law expected from these models is
 $\Delta\Omega \propto (\frac{T_{\rm eff}}{5500 K})^2$ in the range 3800 -- 6000 K and $\Delta\Omega \propto (\frac{T_{\rm eff}}{5500 \rm K})^{20}$ above 6000 K \citep{2011AN....332..933K}.
\par \citet{2013A&A...560A...4R}  analyze the Quarter 3 (Q3) long cadence  photometric data collected by the  Kepler mission \citep{2010Sci...327..977B} and search for SDR in a sample of about 20000 main--sequence late--type stars. 
  In contrast with  \citet{2005MNRAS.357L...1B} and  \citet{2007AN....328.1030C}, they find that SDR exhibits  a weak dependence on temperature in the range 3800 -- 6000  K and that, above 6000 K, it steeply increases towards higher temperatures.
\citet{2013A&A...560A...4R} claim that their results, though quite scattered, are in better agreement with the model of \citet{2011AN....332..933K}. Note however that  these authors have no information on the ages of their targets.  This lack of information  makes  it difficult to compare their results with those of \citet{2005MNRAS.357L...1B} and  \citet{2007AN....328.1030C}. Indeed, the targets investigated by\citet{2005MNRAS.357L...1B} and  \citet{2007AN....328.1030C} comprise  pre-main-sequence stars and stars recently settled in ZAMS while the sample of  \citet{2013A&A...560A...4R} includes field stars of different ages. Moreover their data do not permit to study SDR as a function of the stellar age. 
\par The present work is focused on a  sample  of 111 late-type stars belonging to 11 nearby young loose associations with known ages  (Table 1).
The ages of these associations span the interval  (4 -- 95 Myr)  and therefore our study concerns the SDR evolution during the transition between the Pre-Main-Sequence and the Main-Sequence phase.  
We search for SDR by analyzing  long-term photometric time-series collected by ASAS \citep[All Sky Automatic Survey,][]{1997AcA....47..467P} and SuperWASP \citep[Wide Angle Search for Planets,][]{2006PASP..118.1407P}. 

The paper is organized in the following way: in Sect. 2 we describe the photometric data; in Sect. 3 we explain the method we employed to measure the quantity  $\Delta\Omega_{\rm phot}$\footnote{The subscript stands for photometric and indicates the kind of data used to measure SDR.}, that is a lower limit for SDR; in Sect. 4 we investigate how $\Delta\Omega_{\rm phot}$ depends on global stellar parameters; in Sect. 5 we discuss the main results of our study and in Sec. 6 the conclusions are drawn.

\section{Data}
\citet{2010A&A...520A..15M,2011A&A...532A..10M} analyze ASAS and SuperWASP time-series for about 300 stars belonging to nearby young associations and measure the rotation period for most of these stars.
In this work we reprocess 99 ASAS time-series and 31 SuperWASP time-series with a new technique (see Sect. 3) for estimating a lower limit $\Delta\Omega_{\rm phot}$ for the rotational shear.
In Table \ref{targets} we list the 111 stars studied here together with their colour indexes and spectral types. 
These targets have been selected among those studied in \citet{2010A&A...520A..15M,2011A&A...532A..10M} by picking up those for which the largest number of photometric measurements is available. For nine of these targets both ASAS and SuperWASP time-series are available.
Note that the ASAS time-series processed here span an interval time longer than those processed in \citet{2010A&A...520A..15M,2011A&A...532A..10M}.
 
\subsection{ASAS photometry}

The ASAS time-series analyzed in the present work are  contained in the ASAS-3 Photometric {\it V}-band catalogue\footnote{http://www.astrouw.edu.pl/asas/}. 
These time-series cover a time-span of about 13 years (from 1997 to 2010) and have a typical photometric error of $0.02 ~ \rm mag$. This error allows the detection of rotational modulation in young solar-like stars where the amplitude modulation ranges between 0.05 and 0.3 mags \citep{2011A&A...532A..10M}. 
In our analysis we use filtered time-series obtained by selecting only the best photometric data.
The time-series sampling   depends on the star coordinates. 
We  divide the ASAS time-series into three groups for which the average interval $t_{\rm mean}$ between two consecutive observations is $\simeq$ 1, 2 and 3 days, respectively.
 
 \subsection{SuperWASP photometry}
The SuperWASP time-series have been downloaded from the first public data archive \citep{2010A2006PASP..118.1407ParchiveA...520L..10B}.
These time-series cover a time span of about four years (from 2004 to 2008).
 Their time coverage is quite irregular and depends on the sky coordinates. 
 The time-series processed in the present work are typically made of two  or three distinct observational seasons covering 60-300 d intervals. 
 During each  season a target is observed every day for a few hours with a sampling rate of about ten minutes in the most favorable sky positions. 
 The SuperWASP data have therefore a better sensitivity to periods  $\le1 \rm d$ than the ASAS data. 
 Our analysis is based on the processed flux measurements obtained through application of the SYSREM algorithm \citep{2005MNRAS.356.1466T}. SuperWASP observations were collected in 2004 without any light filter. Starting from 2006, they were collected through a wide band filter in the range 400 -- 700 nm. Owing to differences in spectral bands with respect to the standard Johnson {\it V} ASAS data, we  analyze the SuperWASP data independently without merging them with ASAS data.

\begin{table*}
\caption{Stellar associations investigated in the present work.} 

\label{association}
\centering
\begin{tabular}{llll}
\hline
Association & N.    & age      & Ref.   \\
                   &         & ( Myr )  &           \\  
\hline
$\epsilon$ Chamaleontis ($\epsilon$ Cha) & 9 & 3-5  & 1, 2\\
$\eta$ Chamaleontis ($\eta$ Cha) & 4& 6-10  & 1, 3, 4\\
TW Hydrae (TWA) & 9 & 8-12 & 5, 6\\
$\beta$ Pictoris ($\beta$ Pic) & 18 & 12-22 &  7, 8\\
Octans (Oct)& 1&20-40 &9, 10\\
Columba (Col)  &14 &20-40 &11, 12\\
$\eta$ Carinae ($\eta$ Car) & 12 & 20-40&  11, 12\\
Tucana-Horologium (Tuc/Hor) & 17 & 20-40&  11, 12\\
Argus (Arg) & 10 & 30-50&  9, 13, 14\\
IC 2391 &3 & 30-50&  13, 14\\
AB Doradus (AB Dor) & 14 & 70-120&  15, 16, 17\\
\hline
\end{tabular}
\tablefoot{For each association we report the number of members for which we estimated $\Delta\Omega_{\rm phot}$, the ages estimate and the age reference.
References:(1) \protect\citet{2013MNRAS.435.1325M}; (2) \protect\citet{2003ApJ...599.1207F}; (3) \protect\citet{1999ApJ...516L..77M}; (4) \protect\citet{2001ASPC..243..591L}; (5) \protect\citet{1999ApJ...512L..63W}; (6)\protect\citet{2006A&A...459..511B}; (7) \protect\citet{2003ApJ...599..342S}; (8) \protect\citet{2007ApJS..169..105M}; (9) \protect\citet{2008hsf2.book..757T}; (10) \protect\citet{2013MNRAS.435.1325M15}; (11) \protect\citet{2001ASPC..244...43T}; (12) \protect\citet{2000ApJ...535..959Z}; 13 \protect\citet{2013MNRAS.431.1005D}; (14)\protect\citet{2004ApJ...614..386B}; (15) \protect\citet{2008ApJ...689.1127M}; (16)\protect\citet{2006ApJ...643.1160L}; (17) \protect\citet{2005ApJ...628L..69L}.}
\end{table*}

\begin{longtab}
\begin{longtable}{lllll}
\caption{\label{targets}List of the targets investigated in the present work}\\
\hline\hline
Target ID & Target Name & Assoc. & $B-V$ & Sp. Type   \\
               &                        &             & ( mag ) &                  \\
\hline\hline
\endfirsthead
\caption{continued.}\\
\hline\hline
Target ID & Target Name & Assoc. & $B-V$ & Sp. Type   \\
               &                        &             & ( mag ) &                  \\  
\hline
\endhead
\hline
\endfoot
ASAS J001353-7441.3 & HIP 1113 & TUC/HOR & 0.74 & G8V\\
ASAS J002409-6211.1 & HIP 1910 & TUC/HOR & 1.4 & M0Ve\\
ASAS J003451-6155.0 & HIP 2729 & TUC/HOR & 1.05 & K4Ve\\
ASAS J004220-7747.7 & TYC  9351-1110-1 & TUC/HOR & 1.06 & K3Ve\\
ASAS J011315-6411.6 & TYC  8852-0264-1 & TUC/HOR & 0.87 & K1V\\
ASAS J015749-2154.1 & HIP 9141 & TUC/HOR & 0.65 & G4V\\
ASAS J020136-1610.0 & BD-16-351 & COL & 1.1 & K5\\
ASAS J020718-5311.9 & HIP 9892 & TUC/HOR & 0.65 & G7V\\
ASAS J024126+0559.3 & HIP 12545 & $\beta$ Pic & 1.21 & K6Ve\\
ASAS J024233-5739.6 & TYC  8497-0995-1 & TUC/HOR & 1.23 & K5Ve\\
ASAS J030942-0934.8 & HIP 14684 & ABDOR & 0.81 & G0\\
ASAS J031909-3507.0 & TYC  7026-0325-1 & TUC/HOR & 1.3 & K7Ve\\
ASAS J033049-4555.9 & TYC  8060-1673-1 & TUC/HOR & 0.95 & K3V\\
ASAS J033156-4359.2 & TYC  7574-0803-1 & TUC/HOR & 1.3 & K6Ve\\
ASAS J034723-0158.3 & HIP 17695 & ABDOR & 1.51 & M3\\
ASAS J045249-1955.0 & TYC  5907-1244-1 & TUC/HOR & 0.87 & XX\\
ASAS J045305-4844.6 & TYC  8080-1206-1 & COL & 0.87 & K2V(e)\\
ASAS J045935+0147.0 & HIP 23200 & $\beta$ Pic & 1.39 & M0.5Ve\\
ASAS J050047-5715.4 & HIP 23309 & $\beta$ Pic & 1.4 & M0Ve\\
ASAS J050230-3959.2 & TYC  7587-0925-1 & ABDOR & 0.88 & K4V\\
ASAS J050651+7221.2 & CD -72 248 & Octans & 0.82 & K0IV\\
ASAS J052845-6526.9 & HIP 25647 & ABDOR & 0.83 & K0V\\
ASAS J052857-3328.3 & TYC  7059-1111-1 & ABDOR & 1.06 & K3Ve\\
ASAS J053705-3932.4 & TYC  7600-0516-1 & TUC/HOR & 0.8 & K1V(e)\\
ASAS J055101-5238.2 & TYC  8520-0032-1 & COL & 0.75 & G9IV\\
ASAS J055329-8156.9 & TYC  9390-0322-1 & CAR & 0.79 & K0V+VI\\
ASAS J055751-3804.1 & TYC  7598-1488-1 & ABDOR & 0.69 & G6V(e)\\
ASAS J060834-3402.9 & TYC  7079-0068-1 & ABDOR & 0.79 & G9Ve\\
ASAS J061828-7202.7 & HIP 29964 & $\beta$ Pic & 1.13 & K4Ve\\
ASAS J062607-4102.9 & TYC  7617-0549-1 & COL & 0.8 & K0V\\
ASAS J062806-4826.9 & TYC  8107-1591-1 & COL & 0.65 & G9V\\
ASAS J063950-6128.7 & HIP 31878 & ABDOR & 1.26 & K7Ve\\
ASAS J064346-7158.6 & HIP 32235 & CAR & 0.7 & G6V\\
ASAS J065623-4646.9 & TYC  8118-0871-1 & COL & 0.78 & K0V(e)\\
ASAS J070030-7941.8 & HIP 33737 & CAR & 0.91 & K2V\\
ASAS J070153-4227.9 & CD -42 2906 & Argus & 0.84 & K1V\\
ASAS J072124-5720.6 & TYC  8559-1016-1 & CAR & 0.64 & K0V+D\\
ASAS J072822-4908.6 & CD -48 2972 & Argus & 0.8 & G8V\\
ASAS J072851-3014.8 & HIP 36349 & ABDOR & 1.44 & M1Ve\\
ASAS J073547-3212.2 & HD61005 & Argus & 0.75 & G8V\\
ASAS J082406-6334.1 & TYC  8929-0927-1 & CAR & 0.63 & G5V\\
ASAS J082844-5205.7 & PMM7422 & IC2391 & 0.69 & G6\\
ASAS J083656-7856.8 & RECX1 & $\eta$ Cha & 1.19 & K4Ve\\
ASAS J084006-5338.1 & PMM1083 & IC2391 & 0.57 & G0\\
ASAS J084200-6218.4 & TYC  8930-0601-1 & CAR & 0.8 & K0V\\
ASAS J084229-7903.9 & RECX4 & $\eta$ Cha & 1.33 & K7\\
ASAS J084300-5354.1 & PMM756 & IC2391 & 0.68 & G9\\
ASAS J084432-7846.6 & RECX10 & $\eta$ Cha & 1.33 & K7\\
ASAS J084708-7859.6 & RECX11 & $\eta$ Cha & 1.0 & K4\\
ASAS J085005-7554.6 & TYC  9395-2139-1 & CAR & 0.76 & G9V\\
ASAS J085156-5355.9 & TYC  8569-3597-1 & CAR & 0.69 & G9V\\
ASAS J085746-5408.6 & TYC  8582-3040-1 & CAR & 0.88 & K2IV\\
ASAS J085752-4941.8 & TYC  8160-0958-1 & CAR & 0.73 & G9V\\
ASAS J085929-5446.8 & TYC  8586-2431-1 & CAR & 0.6 & G5IV\\
ASAS J092335-6111.6 & HIP 46063 & CAR & 0.86 & K1V(e)\\
ASAS J092854-4101.3 & TYC  7695-0335-1 & Argus & 0.67 & K3V\\
ASAS J094247-7239.8 & TYC  9217-0641-1 & Argus & 0.65 & K1V\\
ASAS J095558-6721.4 & HD309851 & Argus & 0.6 & G1V\\
ASAS J101315-5230.9 & TWA 21 & TWA & 1.0 & K3V3\\
ASAS J105351-7002.3 & CP -69 1432 & Argus & 0.62 & G2V\\
ASAS J105749-6914.0 & CP -68 1388 & $\epsilon$ Cha & 0.86 & K1V(e)\\
ASAS J110914-3001.7 & TWA 2 & TWA & 1.48 & M2Ve\\
ASAS J112105-3845.3 & TWA 12 & TWA & 1.53 & M1V\\
ASAS J112117-3446.8 & TWA 13A & TWA & 1.42 & M1Ve\\
ASAS J112205-2446.7 & TWA 4 & TWA & 1.17 & K5V\\
ASAS J115942-7601.4 & HIP 58490 & $\epsilon$ Cha & 1.11 & K4Ve\\
ASAS J120139-7859.3 & HD104467 & $\epsilon$ Cha & 0.63 & G5Ve\\
ASAS J120204-7853.1 & GSC 09420-00948 & $\epsilon$ Cha & 1.0 & K7e\\
ASAS J121138-7110.6 & HD105923 & $\epsilon$ Cha & 0.73 & G8V\\
ASAS J121531-3948.7 & TWA 25 & TWA & 1.41 & M1Ve\\
ASAS J122023-7407.7 & GSC 9239-1572 & $\epsilon$ Cha & 0.97 & K7Ve\\
ASAS J122034-7539.5 & CD -74 673 & Argus & 1.02 & K3Ve\\
ASAS J122105-7116.9 & GSC 9235-1702 & $\epsilon$ Cha & 1.2 & K7V\\
ASAS J123921-7502.7 & CD -74 712 & $\epsilon$ Cha & 0.97 & K3Ve\\
ASAS J125826-7028.8 & CD -69 1055 & $\epsilon$ Cha & 0.83 & K0Ve\\
ASAS J134913-7549.8 & CD -75 652 & Argus & 0.68 & G1V\\
ASAS J153857-5742.5 & HIP 76629 & $\beta$ Pic & 0.81 & K0V\\
ASAS J171726-6657.1 & HIP 84586 & $\beta$ Pic & 0.76 & G5IV\\
ASAS J181411-3247.5 & V4046-Sgr & $\beta$ Pic & 0.95 & K5\\
ASAS J181952-2916.5 & HIP 89829 & $\beta$ Pic & 0.69 & G1V\\
ASAS J184653-6210.6 & TYC  9073-0762-1 & $\beta$ Pic & 1.46 & M1Ve\\
ASAS J185306-5010.8 & HIP 92680 & $\beta$ Pic & 0.77 & K8Ve\\
ASAS J200724-5147.5 & CD -52 9381 & Argus & 1.24 & K6Ve\\
ASAS J204510-3120.4 & HIP 102409 & $\beta$ Pic & 1.49 & M1Ve\\
ASAS J205603-1710.9 & TYC  6349-0200-1 & $\beta$ Pic & 1.22 & K6Ve+M\\
ASAS J212050-5302.0 & HIP 105388 & TUC/HOR & 0.72 & G7V\\
ASAS J214430-6058.6 & HIP 107345 & TUC/HOR & 1.41 & M0Ve\\
ASAS J232749-8613.3 & TYC  9529-0340-1 & TUC/HOR & 0.6 & 99\\
ASAS J233231-1215.9 & TYC  5832-0666-1 & $\beta$ Pic & 1.43 & M0Ve\\
ASAS J234154-3558.7 & HIP 116910 & ABDOR & 0.71 & G8V\\
SWASP1  J002334.66+201428.6 & TYC  1186-706-1 & $\beta$ Pic & 1.4 & K7Ve\\
SWASP1  J021055.38-460358.6 & TYC  8042-1050-1 & ABDOR & 0.91 & K3IVe\\
SWASP1  J022729.25+305824.6 & HIP 11437 & $\beta$ Pic & 1.21 & K8\\
SWASP1  J033120.80-303058.7 & HIP 16413 & COL & 0.64 & G7IV\\
SWASP1  J041422.57-381901.5 & HIP 19775 & COL & 0.58 & G3V\\
SWASP1  J042148.68-431732.5 & TYC  7584-1630-1 & COL & 0.69 & G7V\\
SWASP1  J043450.78-354721.2 & TYC  7044-0535-1 & COL & 0.84 & K1Ve\\
SWASP1  J045153.54-464713.3 & TYC  8077-0657-1 & COL & 0.69 & G5V\\
SWASP1  J050649.47-213503.7 & BD-21-1074 & $\beta$ Pic & 1.52 & M2\\
SWASP1  J051829.04-300132.0 & TYC  7048-1453-1 & TUC/HOR & 1.27 & K4Ve\\
SWASP1  J052855.09-453458.3 & TYC  8086-0954-1 & COL & 0.86 & K1V\\
SWASP1  J053504.11-341751.9 & TYC  7064-0839-1 & ABDOR & 1.08 & K4Ve\\
SWASP1  J054516.24-383649.1 & TYC  7597-0833-1 & COL & 0.68 & G9V\\
SWASP1  J055021.43-291520.7 & TYC  6502-1188-1 & COL & 0.66 & K0V(e)\\
SWASP1  J064118.50-382036.1 & TYC  7627-2190-1 & ABDOR & 1.19 & K2e\\
SWASP1  J101828.70-315002.8 & TWA 6 & TWA & 1.31 & M0Ve\\
SWASP1  J113241.23-265200.7 & TWA 8 & TWA & 1.46 & M3Ve\\
SWASP1  J114824.21-372849.2 & TWA 9 & TWA & 1.26 & K5V\\
SWASP1  J191144.66-260408.5 & TYC  6878-0195-1 & $\beta$ Pic & 1.05 & K4V(e)\\
SWASP1  J224457.83-331500.6 & HIP 112312 & $\beta$ Pic & 1.48 & M4IVe\\
SWASP1  J231152.05-450810.6 & HIP 114530 & ABDOR & 0.8 & G8V\\
\end{longtable}
\tablefoot{The color indexes and spectral types are those reported in \protect\citet{2010A&A...520A..15M,2011A&A...532A..10M}. (See the reference therein for details).}
\end{longtab}

\begin{longtab}
\begin{longtable}{lllllllll}
\caption{\label{results}Results}\\
\hline\hline
Star Id. & $<P_{\rm rot}>$ & $\Omega_{\rm min}$ & $\Omega_{\rm max}$ & $\Delta\Omega_{\rm phot}$&  $\alpha_{\rm phot}$& TS& $N_{\rm seg}$& Qual. \\
\hline
        & ( d ) &( $\rm rad~d^{-1}$ ) &  ( $\rm rad~d^{-1}$ )  &  ( $\rm rad~d^{-1}$ ) & ( $\rm rad ~d^{-1}$ ) & & & \\\hline
\endfirsthead
\caption{continued.}\\
\hline\hline
Star Id. & $<P_{\rm rot}>$ & $\Omega_{\rm min}$ & $\Omega_{\rm max}$ & $\Delta\Omega_{\rm phot}$&  $\alpha_{\rm phot}$& TS & $N_{\rm seg}$ & Qual. \\
\hline
        & ( d ) &( $\rm rad~d^{-1}$ ) &  ( $\rm rad~d^{-1}$ )  &  ( $\rm rad~d^{-1}$ ) & ( $\rm rad ~d^{-1}$ ) & & &  \\\hline
\endhead
\hline
\endfoot
ASAS J001353-7441.3 & 3.67 & 1.653 $\pm$ 0.013 & 1.799 $\pm$ 0.016 & 0.146 $\pm$ 0.021 & 0.081 $\pm$ 0.012 & as &	141& B\\
ASAS J002409-6211.1 & 1.75 & 3.558 $\pm$ 0.008 & 3.627 $\pm$ 0.012 & 0.068 $\pm$ 0.014 & 0.019 $\pm$ 0.004 & as &	64& C\\
ASAS J003451-6155.0 & 0.38 & 16.638 $\pm$ 0.006 & 16.695 $\pm$ 0.008 & 0.057 $\pm$ 0.01 & 0.003 $\pm$ 0.001 & as &	79& C\\
ASAS J004220-7747.7 & 2.57 & 2.422 $\pm$ 0.012 & 2.47 $\pm$ 0.01 & 0.048 $\pm$ 0.016 & 0.019 $\pm$ 0.006 & as &	175& B\\
ASAS J011315-6411.6 & 1.26 & 4.956 $\pm$ 0.01 & 5.001 $\pm$ 0.005 & 0.045 $\pm$ 0.011 & 0.009 $\pm$ 0.002 & as &	133& B\\
ASAS J015749-2154.1 & 3.05 & 2.033 $\pm$ 0.016 & 2.119 $\pm$ 0.008 & 0.086 $\pm$ 0.018 & 0.041 $\pm$ 0.009 & sw &	113& B\\
ASAS J020136-1610.0 & 3.21 & 1.909 $\pm$ 0.013 & 1.985 $\pm$ 0.012 & 0.076 $\pm$ 0.018 & 0.038 $\pm$ 0.009 & as &	72& C\\
ASAS J020718-5311.9 & 2.33 & 2.61 $\pm$ 0.012 & 2.843 $\pm$ 0.007 & 0.233 $\pm$ 0.014 & 0.082 $\pm$ 0.005 & as &	63& C\\
ASAS J024126+0559.3 & 4.83 & 1.283 $\pm$ 0.008 & 1.318 $\pm$ 0.01 & 0.035 $\pm$ 0.013 & 0.027 $\pm$ 0.01 & as &	68& C\\
ASAS J024233-5739.6 & 7.4 & 0.816 $\pm$ 0.007 & 0.887 $\pm$ 0.011 & 0.071 $\pm$ 0.013 & 0.08 $\pm$ 0.015 & as &	255& A\\
ASAS J030942-0934.8 & 5.47 & 1.109 $\pm$ 0.009 & 1.177 $\pm$ 0.008 & 0.068 $\pm$ 0.012 & 0.058 $\pm$ 0.01 & as &	45& C\\
ASAS J031909-3507.0 & 8.52 & 0.707 $\pm$ 0.012 & 0.768 $\pm$ 0.011 & 0.061 $\pm$ 0.016 & 0.079 $\pm$ 0.021 & as &	115& B\\
ASAS J033049-4555.9 & 3.8 & 1.591 $\pm$ 0.014 & 1.713 $\pm$ 0.009 & 0.122 $\pm$ 0.017 & 0.071 $\pm$ 0.01 & as &	66& C\\
ASAS J033156-4359.2 & 2.93 & 2.113 $\pm$ 0.009 & 2.182 $\pm$ 0.008 & 0.07 $\pm$ 0.012 & 0.032 $\pm$ 0.005 & as &	89& C\\
ASAS J034723-0158.3 & 3.86 & 1.614 $\pm$ 0.01 & 1.672 $\pm$ 0.009 & 0.057 $\pm$ 0.013 & 0.034 $\pm$ 0.008 & as &	14& C\\
ASAS J045249-1955.0 & 5.21 & 1.178 $\pm$ 0.01 & 1.238 $\pm$ 0.01 & 0.06 $\pm$ 0.014 & 0.049 $\pm$ 0.011 & as &	129& B\\
ASAS J045305-4844.6 & 4.59 & 1.344 $\pm$ 0.007 & 1.403 $\pm$ 0.01 & 0.059 $\pm$ 0.012 & 0.042 $\pm$ 0.009 & as &	153& B\\
ASAS J045935+0147.0 & 4.42 & 1.4 $\pm$ 0.009 & 1.447 $\pm$ 0.005 & 0.047 $\pm$ 0.01 & 0.033 $\pm$ 0.007 & as &	100& B\\
ASAS J050047-5715.4 & 8.74 & 0.694 $\pm$ 0.011 & 0.743 $\pm$ 0.007 & 0.05 $\pm$ 0.013 & 0.067 $\pm$ 0.017 & as &	173& B\\
ASAS J050230-3959.2 & 6.58 & 0.92 $\pm$ 0.007 & 0.993 $\pm$ 0.015 & 0.073 $\pm$ 0.017 & 0.073 $\pm$ 0.017 & as &	247& A\\
ASAS J050651-7221.2 & 0.24 & 26.637 $\pm$ 0.012 & 26.738 $\pm$ 0.012 & 0.101 $\pm$ 0.017 & 0.004 $\pm$ 0.001 & as &	44& C\\
ASAS J052845-6526.9 & 0.51 & 12.189 $\pm$ 0.015 & 12.242 $\pm$ 0.013 & 0.053 $\pm$ 0.02 & 0.004 $\pm$ 0.002 & as &	273& A\\
ASAS J052857-3328.3 & 0.69 & 9.005 $\pm$ 0.008 & 9.075 $\pm$ 0.01 & 0.07 $\pm$ 0.013 & 0.008 $\pm$ 0.001 & as &	43& C\\
ASAS J053705-3932.4 & 2.46 & 2.44 $\pm$ 0.007 & 2.622 $\pm$ 0.009 & 0.182 $\pm$ 0.011 & 0.069 $\pm$ 0.004 & sw &	390& A\\
ASAS J055101-5238.2 & 1.2 & 5.182 $\pm$ 0.013 & 5.263 $\pm$ 0.014 & 0.081 $\pm$ 0.021 & 0.015 $\pm$ 0.004 & as &	122& B\\
ASAS J055329-8156.9 & 1.86 & 3.313 $\pm$ 0.012 & 3.438 $\pm$ 0.02 & 0.125 $\pm$ 0.024 & 0.036 $\pm$ 0.007 & as &	406& A\\
ASAS J055751-3804.1 & 0.79 & 7.911 $\pm$ 0.009 & 8.054 $\pm$ 0.011 & 0.143 $\pm$ 0.014 & 0.018 $\pm$ 0.002 & as &	126& B\\
ASAS J060834-3402.9 & 3.4 & 1.797 $\pm$ 0.01 & 1.897 $\pm$ 0.013 & 0.1 $\pm$ 0.018 & 0.053 $\pm$ 0.01 & as &	119& B\\
ASAS J061828-7202.7 & 2.67 & 2.333 $\pm$ 0.013 & 2.378 $\pm$ 0.007 & 0.045 $\pm$ 0.016 & 0.019 $\pm$ 0.007 & as &	174& B\\
ASAS J062607-4102.9 & 4.18 & 1.451 $\pm$ 0.005 & 1.558 $\pm$ 0.009 & 0.106 $\pm$ 0.015 & 0.068 $\pm$ 0.01 & sw &	147& B\\
ASAS J062806-4826.9 & 1.29 & 4.823 $\pm$ 0.015 & 4.901 $\pm$ 0.015 & 0.078 $\pm$ 0.021 & 0.016 $\pm$ 0.004 & as &	80& C\\
ASAS J063950-6128.7 & 9.12 & 0.666 $\pm$ 0.017 & 0.711 $\pm$ 0.01 & 0.045 $\pm$ 0.021 & 0.064 $\pm$ 0.03 & as &	32& C\\
ASAS J064346-7158.6 & 3.89 & 1.541 $\pm$ 0.007 & 1.669 $\pm$ 0.011 & 0.128 $\pm$ 0.013 & 0.077 $\pm$ 0.008 & as &	155& B\\
ASAS J065623-4646.9 & 4.45 & 1.348 $\pm$ 0.02 & 1.473 $\pm$ 0.012 & 0.124 $\pm$ 0.025 & 0.084 $\pm$ 0.017 & as &	156& B\\
ASAS J070030-7941.8 & 5.11 & 1.189 $\pm$ 0.008 & 1.269 $\pm$ 0.01 & 0.08 $\pm$ 0.013 & 0.063 $\pm$ 0.01 & as &	413& A\\
ASAS J070153-4227.9 & 3.99 & 1.526 $\pm$ 0.015 & 1.641 $\pm$ 0.016 & 0.114 $\pm$ 0.022 & 0.07 $\pm$ 0.014 & as &	230& A\\
ASAS J072124-5720.6 & 4.67 & 1.285 $\pm$ 0.015 & 1.399 $\pm$ 0.01 & 0.114 $\pm$ 0.018 & 0.082 $\pm$ 0.013 & as &	129& B\\
ASAS J072822-4908.6 & 1.03 & 5.996 $\pm$ 0.012 & 6.157 $\pm$ 0.012 & 0.161 $\pm$ 0.017 & 0.026 $\pm$ 0.003 & as &	101& B\\
ASAS J072851-3014.8 & 1.64 & 3.802 $\pm$ 0.015 & 3.849 $\pm$ 0.012 & 0.047 $\pm$ 0.019 & 0.012 $\pm$ 0.005 & as &	226& A\\
ASAS J073547-3212.2 & 5.06 & 1.199 $\pm$ 0.019 & 1.286 $\pm$ 0.017 & 0.087 $\pm$ 0.025 & 0.068 $\pm$ 0.02 & as &	51& C\\
ASAS J082406-6334.1 & 0.79 & 7.842 $\pm$ 0.015 & 8.03 $\pm$ 0.019 & 0.188 $\pm$ 0.025 & 0.023 $\pm$ 0.003 & as &	204& A\\
ASAS J082844-5205.7 & 1.51 & 4.116 $\pm$ 0.01 & 4.21 $\pm$ 0.012 & 0.094 $\pm$ 0.016 & 0.022 $\pm$ 0.004 & as &	169& B\\
ASAS J083656-7856.8 & 4.45 & 1.384 $\pm$ 0.007 & 1.454 $\pm$ 0.008 & 0.07 $\pm$ 0.011 & 0.048 $\pm$ 0.008 & as &	233& A\\
ASAS J084006-5338.1 & 1.34 & 4.662 $\pm$ 0.007 & 4.746 $\pm$ 0.01 & 0.085 $\pm$ 0.012 & 0.018 $\pm$ 0.003 & as &	89& C\\
ASAS J084200-6218.4 & 1.22 & 5.085 $\pm$ 0.011 & 5.183 $\pm$ 0.005 & 0.097 $\pm$ 0.012 & 0.019 $\pm$ 0.002 & as &	140& B\\
ASAS J084229-7903.9 & 7.14 & 0.838 $\pm$ 0.014 & 0.924 $\pm$ 0.015 & 0.087 $\pm$ 0.021 & 0.094 $\pm$ 0.023 & as &	204& A\\
ASAS J084300-5354.1 & 3.15 & 1.957 $\pm$ 0.012 & 2.028 $\pm$ 0.011 & 0.072 $\pm$ 0.017 & 0.035 $\pm$ 0.008 & as &	131& B\\
ASAS J084432-7846.6 & 20.26 & 0.265 $\pm$ 0.013 & 0.336 $\pm$ 0.017 & 0.07 $\pm$ 0.022 & 0.21 $\pm$ 0.067 & as &	74& C\\
ASAS J084708-7859.6 & 4.84 & 1.245 $\pm$ 0.014 & 1.355 $\pm$ 0.01 & 0.111 $\pm$ 0.017 & 0.082 $\pm$ 0.013 & as &	104& B\\
ASAS J085005-7554.6 & 1.15 & 5.433 $\pm$ 0.017 & 5.55 $\pm$ 0.016 & 0.117 $\pm$ 0.023 & 0.021 $\pm$ 0.004 & as &	183& B\\
ASAS J085156-5355.9 & 1.91 & 3.172 $\pm$ 0.018 & 3.397 $\pm$ 0.014 & 0.226 $\pm$ 0.023 & 0.066 $\pm$ 0.007 & as &	81& C\\
ASAS J085746-5408.6 & 1.94 & 3.202 $\pm$ 0.022 & 3.286 $\pm$ 0.012 & 0.084 $\pm$ 0.025 & 0.025 $\pm$ 0.007 & as &	70& C\\
ASAS J085752-4941.8 & 2.03 & 3.005 $\pm$ 0.002 & 3.149 $\pm$ 0.009 & 0.143 $\pm$ 0.009 & 0.045 $\pm$ 0.003 & as &	47& C\\
ASAS J085929-5446.8 & 0.44 & 14.264 $\pm$ 0.011 & 14.375 $\pm$ 0.006 & 0.111 $\pm$ 0.013 & 0.008 $\pm$ 0.001 & as &	162& B\\
ASAS J092335-6111.6 & 3.89 & 1.582 $\pm$ 0.018 & 1.648 $\pm$ 0.01 & 0.066 $\pm$ 0.021 & 0.04 $\pm$ 0.013 & as &	244& A\\
ASAS J092854-4101.3 & 0.39 & 16.011 $\pm$ 0.006 & 16.076 $\pm$ 0.008 & 0.065 $\pm$ 0.01 & 0.004 $\pm$ 0.001 & as &	69& C\\
ASAS J094247-7239.8 & 2.31 & 2.682 $\pm$ 0.02 & 2.751 $\pm$ 0.01 & 0.07 $\pm$ 0.022 & 0.025 $\pm$ 0.008 & as &	143& B\\
ASAS J095558-6721.4 & 1.83 & 3.414 $\pm$ 0.016 & 3.499 $\pm$ 0.008 & 0.085 $\pm$ 0.019 & 0.024 $\pm$ 0.005 & as &	133& B\\
ASAS J101315-5230.9 & 4.4 & 1.388 $\pm$ 0.008 & 1.482 $\pm$ 0.014 & 0.094 $\pm$ 0.016 & 0.063 $\pm$ 0.011 & as &	164& B\\
ASAS J105351-7002.3 & 1.03 & 6.056 $\pm$ 0.014 & 6.157 $\pm$ 0.007 & 0.102 $\pm$ 0.017 & 0.017 $\pm$ 0.003 & as &	32& C\\
ASAS J105749-6914.0 & 3.58 & 1.728 $\pm$ 0.013 & 1.781 $\pm$ 0.006 & 0.053 $\pm$ 0.014 & 0.03 $\pm$ 0.008 & as &	184& B\\
ASAS J110914-3001.7 & 4.85 & 1.255 $\pm$ 0.018 & 1.341 $\pm$ 0.014 & 0.086 $\pm$ 0.023 & 0.064 $\pm$ 0.017 & as &	102& B\\
ASAS J112105-3845.3 & 3.3 & 1.882 $\pm$ 0.011 & 1.921 $\pm$ 0.009 & 0.039 $\pm$ 0.014 & 0.02 $\pm$ 0.007 & as &	138& B\\
ASAS J112117-3446.8 & 5.44 & 1.108 $\pm$ 0.007 & 1.187 $\pm$ 0.003 & 0.079 $\pm$ 0.008 & 0.067 $\pm$ 0.007 & sw &	352& A\\
ASAS J112205-2446.7 & 14.3 & 0.413 $\pm$ 0.015 & 0.469 $\pm$ 0.014 & 0.056 $\pm$ 0.021 & 0.119 $\pm$ 0.045 & as &	108& B\\
ASAS J115942-7601.4 & 8.04 & 0.756 $\pm$ 0.011 & 0.808 $\pm$ 0.011 & 0.052 $\pm$ 0.016 & 0.065 $\pm$ 0.02 & as &	87& C\\
ASAS J120139-7859.3 & 4.38 & 1.368 $\pm$ 0.012 & 1.492 $\pm$ 0.018 & 0.124 $\pm$ 0.022 & 0.083 $\pm$ 0.015 & as &	34& C\\
ASAS J120204-7853.1 & 4.44 & 1.393 $\pm$ 0.005 & 1.434 $\pm$ 0.023 & 0.041 $\pm$ 0.024 & 0.028 $\pm$ 0.016 & as &	161& B\\
ASAS J121138-7110.6 & 5.13 & 1.182 $\pm$ 0.008 & 1.28 $\pm$ 0.009 & 0.098 $\pm$ 0.012 & 0.076 $\pm$ 0.009 & as &	107& B\\
ASAS J121531-3948.7 & 5.06 & 1.216 $\pm$ 0.017 & 1.263 $\pm$ 0.008 & 0.047 $\pm$ 0.019 & 0.037 $\pm$ 0.015 & as &	224& A\\
ASAS J122023-7407.7 & 1.54 & 4.052 $\pm$ 0.013 & 4.129 $\pm$ 0.017 & 0.077 $\pm$ 0.022 & 0.019 $\pm$ 0.005 & as &	167& B\\
ASAS J122034-7539.5 & 3.49 & 1.756 $\pm$ 0.007 & 1.844 $\pm$ 0.015 & 0.088 $\pm$ 0.017 & 0.048 $\pm$ 0.009 & as &	178& B\\
ASAS J122105-7116.9 & 6.86 & 0.892 $\pm$ 0.005 & 0.942 $\pm$ 0.011 & 0.05 $\pm$ 0.012 & 0.053 $\pm$ 0.013 & as &	113& B\\
ASAS J123921-7502.7 & 3.99 & 1.548 $\pm$ 0.008 & 1.598 $\pm$ 0.01 & 0.05 $\pm$ 0.013 & 0.031 $\pm$ 0.008 & as &	331& A\\
ASAS J125826-7028.8 & 2.0 & 3.13 $\pm$ 0.013 & 3.178 $\pm$ 0.011 & 0.048 $\pm$ 0.017 & 0.015 $\pm$ 0.005 & as &	49& C\\
ASAS J134913-7549.8 & 2.29 & 2.703 $\pm$ 0.014 & 2.802 $\pm$ 0.009 & 0.1 $\pm$ 0.017 & 0.036 $\pm$ 0.006 & as &	171& B\\
ASAS J153857-5742.5 & 4.3 & 1.412 $\pm$ 0.014 & 1.517 $\pm$ 0.011 & 0.105 $\pm$ 0.018 & 0.069 $\pm$ 0.012 & as &	210& A\\
ASAS J171726-6657.1 & 1.68 & 3.691 $\pm$ 0.016 & 3.809 $\pm$ 0.015 & 0.119 $\pm$ 0.023 & 0.031 $\pm$ 0.006 & as &	168& B\\
ASAS J181411-3247.5 & 2.42 & 2.553 $\pm$ 0.003 & 2.65 $\pm$ 0.007 & 0.097 $\pm$ 0.008 & 0.037 $\pm$ 0.003 & as &	98& C\\
ASAS J181952-2916.5 & 0.57 & 10.969 $\pm$ 0.013 & 11.066 $\pm$ 0.013 & 0.097 $\pm$ 0.018 & 0.009 $\pm$ 0.002 & as &	194& B\\
ASAS J184653-6210.6 & 5.37 & 1.146 $\pm$ 0.009 & 1.199 $\pm$ 0.012 & 0.053 $\pm$ 0.015 & 0.044 $\pm$ 0.012 & as &	88& C\\
ASAS J185306-5010.8 & 0.94 & 6.625 $\pm$ 0.012 & 6.696 $\pm$ 0.014 & 0.072 $\pm$ 0.018 & 0.011 $\pm$ 0.003 & as &	39& C\\
ASAS J200724-5147.5 & 0.84 & 7.484 $\pm$ 0.011 & 7.521 $\pm$ 0.005 & 0.036 $\pm$ 0.012 & 0.005 $\pm$ 0.002 & as &	62& C\\
ASAS J204510-3120.4 & 4.84 & 1.269 $\pm$ 0.009 & 1.334 $\pm$ 0.012 & 0.065 $\pm$ 0.015 & 0.049 $\pm$ 0.011 & as &	31& C\\
ASAS J205603-1710.9 & 3.4 & 1.828 $\pm$ 0.008 & 1.87 $\pm$ 0.015 & 0.042 $\pm$ 0.017 & 0.023 $\pm$ 0.009 & as &	35& C\\
ASAS J212050-5302.0 & 3.43 & 1.711 $\pm$ 0.018 & 1.906 $\pm$ 0.01 & 0.195 $\pm$ 0.021 & 0.102 $\pm$ 0.011 & as &	108& B\\
ASAS J214430-6058.6 & 4.55 & 1.342 $\pm$ 0.006 & 1.428 $\pm$ 0.017 & 0.086 $\pm$ 0.018 & 0.06 $\pm$ 0.013 & as &	52& C\\
ASAS J232749-8613.3 & 0.7 & 8.882 $\pm$ 0.011 & 9.05 $\pm$ 0.017 & 0.168 $\pm$ 0.02 & 0.019 $\pm$ 0.002 & as &	226& A\\
ASAS J233231-1215.9 & 5.69 & 1.086 $\pm$ 0.007 & 1.117 $\pm$ 0.009 & 0.031 $\pm$ 0.011 & 0.028 $\pm$ 0.01 & as &	42& C\\
ASAS J234154-3558.7 & 1.79 & 3.419 $\pm$ 0.017 & 3.566 $\pm$ 0.005 & 0.147 $\pm$ 0.018 & 0.041 $\pm$ 0.005 & sw &	242& A\\
SWASP1 J002334.66+201428.6 & 7.88 & 0.753 $\pm$ 0.02 & 0.841 $\pm$ 0.009 & 0.088 $\pm$ 0.025 & 0.105 $\pm$ 0.03 & sw &	401& A\\
SWASP1 J021055.38-460358.6 & 1.12 & 5.586 $\pm$ 0.008 & 5.664 $\pm$ 0.005 & 0.077 $\pm$ 0.01 & 0.014 $\pm$ 0.002 & sw &	131& B\\
SWASP1 J022729.25+305824.6 & 12.43 & 0.365 $\pm$ 0.01 & 0.521 $\pm$ 0.013 & 0.156 $\pm$ 0.018 & 0.299 $\pm$ 0.035 & sw &	142& B\\
SWASP1 J033120.80-303058.7 & 1.85 & 3.355 $\pm$ 0.006 & 3.418 $\pm$ 0.006 & 0.063 $\pm$ 0.008 & 0.018 $\pm$ 0.002 & sw &	192& B\\
SWASP1 J041422.57-381901.5 & 1.69 & 3.672 $\pm$ 0.013 & 3.769 $\pm$ 0.012 & 0.097 $\pm$ 0.018 & 0.026 $\pm$ 0.005 & sw &	147& B\\
SWASP1 J042148.68-431732.5 & 1.95 & 3.158 $\pm$ 0.015 & 3.284 $\pm$ 0.005 & 0.126 $\pm$ 0.016 & 0.038 $\pm$ 0.005 & sw &	243& A\\
SWASP1 J043450.78-354721.2 & 2.3 & 2.65 $\pm$ 0.017 & 2.773 $\pm$ 0.003 & 0.123 $\pm$ 0.023 & 0.044 $\pm$ 0.008 & sw &	298& A\\
SWASP1 J045153.54-464713.3 & 2.84 & 2.161 $\pm$ 0.007 & 2.239 $\pm$ 0.012 & 0.078 $\pm$ 0.014 & 0.035 $\pm$ 0.006 & sw &	223& A\\
SWASP1 J050649.47-213503.7 & 13.39 & 0.453 $\pm$ 0.007 & 0.508 $\pm$ 0.01 & 0.055 $\pm$ 0.012 & 0.108 $\pm$ 0.024 & sw &	227& A\\
SWASP1 J051829.04-300132.0 & 1.7 & 3.686 $\pm$ 0.008 & 3.725 $\pm$ 0.007 & 0.039 $\pm$ 0.011 & 0.01 $\pm$ 0.003 & sw &	223& A\\
SWASP1 J052855.09-453458.3 & 4.68 & 1.298 $\pm$ 0.009 & 1.397 $\pm$ 0.004 & 0.099 $\pm$ 0.01 & 0.071 $\pm$ 0.007 & sw &	277& A\\
SWASP1 J053504.11-341751.9 & 1.15 & 5.472 $\pm$ 0.003 & 5.495 $\pm$ 0.004 & 0.023 $\pm$ 0.005 & 0.004 $\pm$ 0.001 & sw &	310& A\\
SWASP1 J054516.24-383649.1 & 1.35 & 4.573 $\pm$ 0.007 & 4.685 $\pm$ 0.003 & 0.112 $\pm$ 0.008 & 0.024 $\pm$ 0.002 & sw &	218& A\\
SWASP1 J055021.43-291520.7 & 3.47 & 1.76 $\pm$ 0.005 & 1.872 $\pm$ 0.006 & 0.112 $\pm$ 0.008 & 0.06 $\pm$ 0.004 & sw &	276& A\\
SWASP1 J064118.50-382036.1 & 0.73 & 8.576 $\pm$ 0.009 & 8.686 $\pm$ 0.008 & 0.109 $\pm$ 0.012 & 0.013 $\pm$ 0.001 & sw &	310& A\\
SWASP1 J101828.70-315002.8 & 0.54 & 11.591 $\pm$ 0.007 & 11.629 $\pm$ 0.006 & 0.038 $\pm$ 0.009 & 0.003 $\pm$ 0.001 & sw &	474& A\\
SWASP1 J113241.23-265200.7 & 4.64 & 1.346 $\pm$ 0.002 & 1.376 $\pm$ 0.004 & 0.03 $\pm$ 0.004 & 0.022 $\pm$ 0.003 & sw &	325& A\\
SWASP1 J114824.21-372849.2 & 5.04 & 1.215 $\pm$ 0.011 & 1.273 $\pm$ 0.013 & 0.058 $\pm$ 0.018 & 0.046 $\pm$ 0.014 & sw &	252& A\\
SWASP1 J191144.66-260408.5 & 5.73 & 1.071 $\pm$ 0.005 & 1.132 $\pm$ 0.006 & 0.061 $\pm$ 0.008 & 0.054 $\pm$ 0.007 & sw &	192& B\\
SWASP1 J224457.83-331500.6 & 2.36 & 2.627 $\pm$ 0.014 & 2.7 $\pm$ 0.007 & 0.074 $\pm$ 0.02 & 0.027 $\pm$ 0.007 & sw &	265& A\\
SWASP1 J231152.05-450810.6 & 5.16 & 1.176 $\pm$ 0.013 & 1.266 $\pm$ 0.009 & 0.09 $\pm$ 0.017 & 0.071 $\pm$ 0.013 & sw &	87& C\\
\end{longtable}
 \tablefoot{For each source, we reported the average rotation period, the minimum and maximum detected angular frequencies, the corresponding $\Delta\Omega_{\rm phot}$ and the relative shear $\alpha_{\rm phot}$. Next to each quantity we also reported the estimated error. In the last three columns we reported a time-series flag indicating wether the results have been inferred with ASAS or Super WASP time-series, the number N of segments in which a significant period has been detected and a quality flag related to N. The flag A is assigned to stars with $N> 200$, the flag B to  stars with N between 100 and 200 and the flag C  to stars with $N<100$ } 
\end{longtab}

\section{The method}
The surface of a solar-like star is covered by magnetically Active Regions (hereafter ARs) associated with dark spots and bright faculae.
If the star is a differential rotator, the ARs will rotate with different frequencies, depending on their latitudes, and will modulate the optical flux coming from the star.
ARs migration in latitude  induces a modulation of the measured photometric period that can be used for estimating the amplitude of SDR. Based on this, the method we employ to measure SDR on a given star consists in the followings steps:

\begin{itemize}
\item{we segment the photometric time-series with a "sliding-window" algorithm;  }
\item{we search for the stellar rotation period in each segment;}
\item{we pick up the maximum and the minimum detected periods and we compute the quantity $\Delta\Omega_{\rm phot}~= ~\Omega_{\rm max}-\Omega_{\rm min}$ where $\Omega_{\rm max}=\frac{2\pi}{P_{\rm min}}$ and  $\Omega_{\rm min}=\frac{2\pi}{P_{\rm max}}$} .
\end{itemize} 
We assume $P_{\rm max}$ and $P_{\rm min}$ are related  to time-series segments where the stellar surface is dominated by high latitude and  low  latitude ARs, respectively (or by the opposite configuration in the case of an anti-solar SDR).
Since the $P_{\rm max}$ and $P_{\rm min}$ thus found do not necessarily sample the whole latitude range, $\Delta\Omega_{\rm phot}$ is obviously a lower limit for the rotational shear and therefore $\Delta\Omega_{\rm phot}\leq\Delta\Omega$. 

\subsection{Time-series segmentation}
Time-series segmentation is very useful in dealing with magnetically active stars.  In fact, the typical light curves of these variables are not very regular because  the flux variation due to  rotational modulation  can be masked by the intrinsic evolution of ARs.

  In the Sun,  ARs evolve on a typical time-scale $\tau_{\rm AR}~\simeq~60$~d \citep{2003A&A...403.1135L}. This time-scale, that is about twice the Sun rotation period, and the simultaneous occurrence of different ARs over the solar-disk could, in principle, destroy the coherence of the rotational signal and prevent the detection of the rotation period. However ARs tend to form, in the Sun, at active longitudes i.e. heliographic longitudes characterized by the frequent, localized emergence of new magnetic flux. 
These AR complexes (ARCs), forming at active longitudes, evolve in a typical time-scale $\tau_{\rm ARC}\simeq~200-250~\rm d$ \citep{2003A&A...403.1135L}. Hence, the rotational signal is approximatively coherent on time intervals of the same order as $\tau_{\rm ARC}$.
\citet{2004A&A...425..707L} analyze the TSI (Total Solar Irradiance) time-series collected by the SoHo satellite and  show  that the  use of period-search algorithms to detect the rotation period
of the Sun  is only possible  if the long term time-series is divided into sub-series with a time extension no longer than $150~\rm d$ in order to limit the effect of the evolution of ARCs.

The  time-scales of ARs and ARCs evolution in late-type stars other than the Sun have been  studied to a limited extent because of the lack of multi-years continuous photometric  time-series.  \citet{1997SoPh..171..191D,1997SoPh..171..211D} estimated the time-scales of ARs evolution in some tens of late-type stars by analyzing long-term time-series of photometric fluxes  in two 0.1 nm passbands centered on the Ca II H and K emission lines. They show that the older and less magnetically active stars tend to be ''AR evolution-dominated''. This means that the evolution of ARs and ARCs take place on time-scales comparable with the stellar rotation period. The light curves of younger and more active stars, conversely, exhibit a  pattern that remains stable for several consecutive rotations. The typical time-scales measured for these stars are $\tau_{\rm AR}~=~50~d$ and $\tau_{\rm ARC}~=~1$~yr. Similar values are reported by \citet{2002AN....323..349H}.
\citet{2003A&A...409.1017M} analyze long-term photometric time-serie collected in the {\it Johnson} V band for six young solar analogues. They find that 2 of these stars are evolution-dominated, while the others have $\tau_{\rm ARC}$ values ranging between 160 d and 500 d, remarkably longer than their rotation periods.

The advantage of using time-series segments is that the data exhibit a more stable pattern of flux variations than the whole time-series, however, they also contain less points. Therefore, the rotation period retrieved by analyzing a time-series segment could have a low statistical significance. A compromise has been found between the length of the segments and the number of points included in the segments by adopting a  length $T=50$ d. In the case of ASAS data, the average number of points per segment range between 16 and 50, depending on the source coordinate.
In the case of Super-WASP data, the average number of points per segment range between 1000 and 3000.
Note that T= 50 d is also the typical $\tau_{\rm AR}$ time-scale found by \citet{1997SoPh..171..191D} for young active stars. In Sec. 3.5  we discuss how the choice of different segments duration can affect $\Delta\Omega_{\rm phot}$ measurements.

\begin{figure*}
\begin{center}
\includegraphics[angle=0]{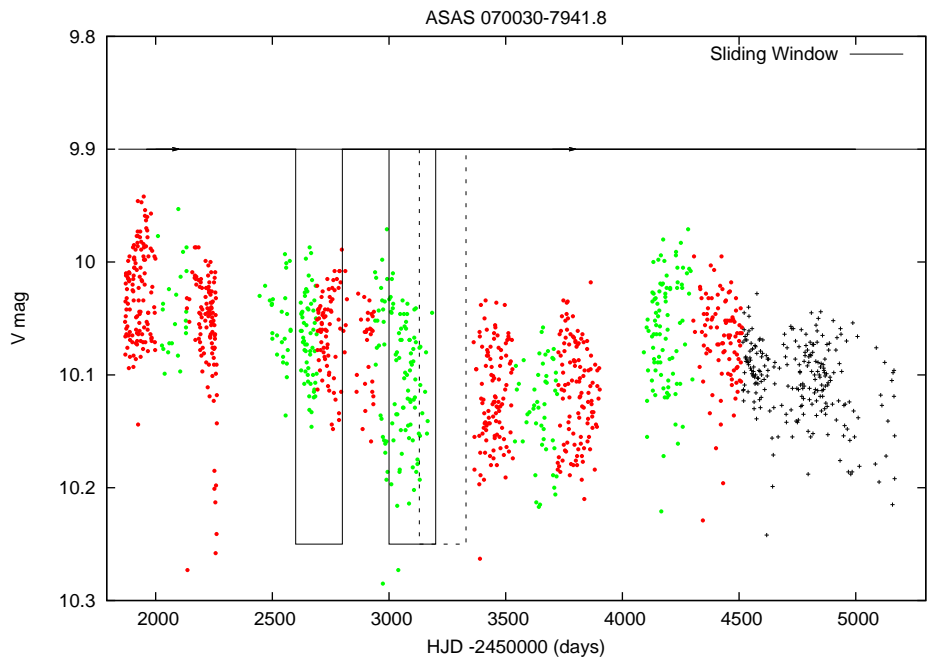}
\caption{ A typical $V$-band time-series collected by the ASAS survey. The red and green bullets mark the segmentation performed by \protect\citet{2010A&A...520A..15M}. The black continuous boxes  depict two of the extracted segments whereas the dotted box indicates one of the sub-series rejected by our segmentation algorithm. The black crosses mark data that have not been processed in \protect\citet{2010A&A...520A..15M}. Note that we employ a 50-day sliding window, whereas in the plot we sketch a 200-day sliding window in order to make the plot more clear.
 }
\label{window}
\end{center}
\end{figure*}

We perform  the  time-series segmentation by using a sliding window algorithm as sketched  in  Fig.  \ref{window}.   This algorithm performs a cycle on the whole time-series. For each observation time $t_i$, a segment is generated by selecting all points in the time interval $[t_i ,t_i + T]$,  where T is the length of the window.
 A segment is rejected if it has less then 10 points or if it is a subset of another segment (this last case  happens when the time-series exhibits gaps larger than the sliding window; see Fig. \ref{window}). 

In Fig. \ref{window} we show the V-Band time-series of the star ASAS J070030-7941.8.
The observations span about nine years. The red and the green bullets  mark the data processed by \citet{2010A&A...520A..15M}. The different colors are used  to  display the segmentation performed in that work. The black crosses mark data collected after February 2008 and that were not processed by \citet{2010A&A...520A..15M}.
 \citet{2010A&A...520A..15M} extracted 10 sub-series. Our sliding window algorithm allows the extraction of more segments.

\subsection{Rotation period and SDR estimate}
We  perform period-search  by means of  the  Lomb-Scargle periodogram, that has been found to be a highly efficient method  in dealing with solar-like variables  \citep{2012MNRAS.421.2774D}. 
For each time-series segment we compute the periodogram and we select the period $P$ with the highest power $Z$.
We build the distribution of the detected periods and we estimate the average stellar rotation period by taking the mode of the distribution.  

\begin{figure*}
\begin{center}
\includegraphics[]{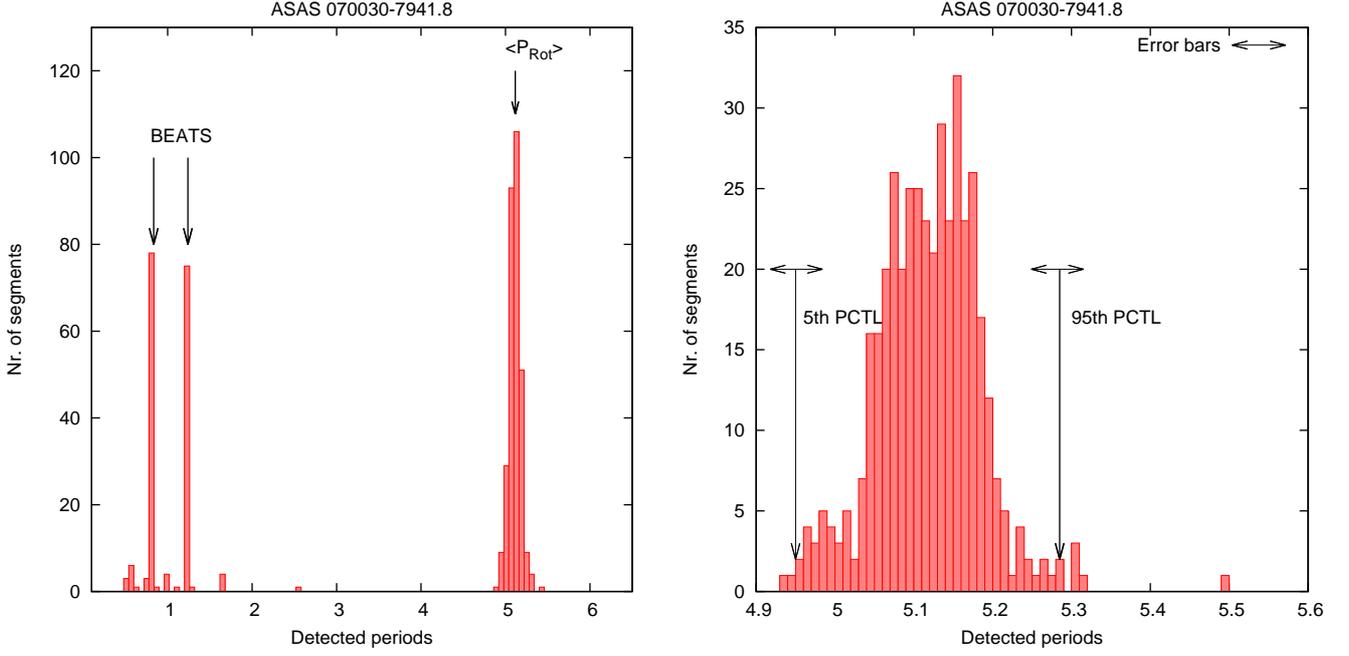}
\caption{{\it Left panel.} Distribution of detected periods for the star ASAS  J070030-7941.8. The black continuous arrow marks the average rotation period $<P_{\rm rot}>$. The black dotted arrows mark the beat periods.  
{\it Right panel.} Period distribution after beats rejections. $P_{\rm min}$ and $P_{\rm max}$ are estimated by taking the 5-th and the 95-th percentile of the distribution in order to avoid  that  $\Delta\Omega_{\rm phot}$ computation is driven by outliers. }
\label{distribution_1}
\end{center}
\end{figure*}

In the left panel of Fig. \ref{distribution_1}, we show the frequency histogram obtained for the source  ASAS  J070030-7941.8. The distribution exhibits a well defined peak at $P_{\rm{rot}}=5.12 ~\rm d$  that is taken as the average stellar rotation period. 
The other peaks  occur at the beat periods $B$ due to the interference between the rotational modulation and the typical ASAS 1-day sampling and are given  by the relationship   $B = P_{\rm rot}/ \left |1 \pm nP_{\rm rot}\right |$ with  $n=1,2,3,\dots$.
 
 Once estimated the average  rotation period, we perform again the period search by rejecting peaks close to the beat frequencies  and selecting, for each segment, the highest significant peak closest to the average rotation period. 
We build  the true period distribution and  pick up $P_{\rm min}$ and $P_{\rm max}$ in order to compute $\Delta\Omega_{\rm phot}$.
In some cases the period distributions show outliers. These outliers could be related to real rotation frequencies or to false positives and could lead to an overestimate of $\Delta\Omega_{\rm phot}$. In order to avoid this, we estimate $P_{\rm min}$ and $P_{\rm max}$ by taking the 5-th and the 95-th percentile of the distribution.
Note that this rejection criterion is very cautious. Indeed, the percentage of measurements distant more than 3 $\sigma$ from the average period is less than 1-2\% for the majority of our targets.
In the right panel of Fig. \ref{distribution_1}, we show the true periods distribution obtained for 
ASAS  J070030-7941.8 after beats rejection. In the picture, we report also the error bars associated with $P_{\rm min}$ and $P_{\rm max}$. In Figs. 3-5 we plot also  the distributions obtained for the stars ASAS J072851-3014.8, ASAS J084229-7903.9 and SWASP1 J101828.70-315002.8 taken as representative examples of our targets.

The total number of segments useful to measure the rotation period is different for each target. This number is a complex function of the time-series sampling and of the stellar rotation period. We divide our data in three quality groups according to the number of segments $N$ in which a  period with $FAP < 0.01$ is measured. The group A includes stars with $N> 200$, the group B  stars with N between 100 and 200 and the group C  stars with $N<100$.

\begin{figure*}
\begin{center}
\includegraphics[]{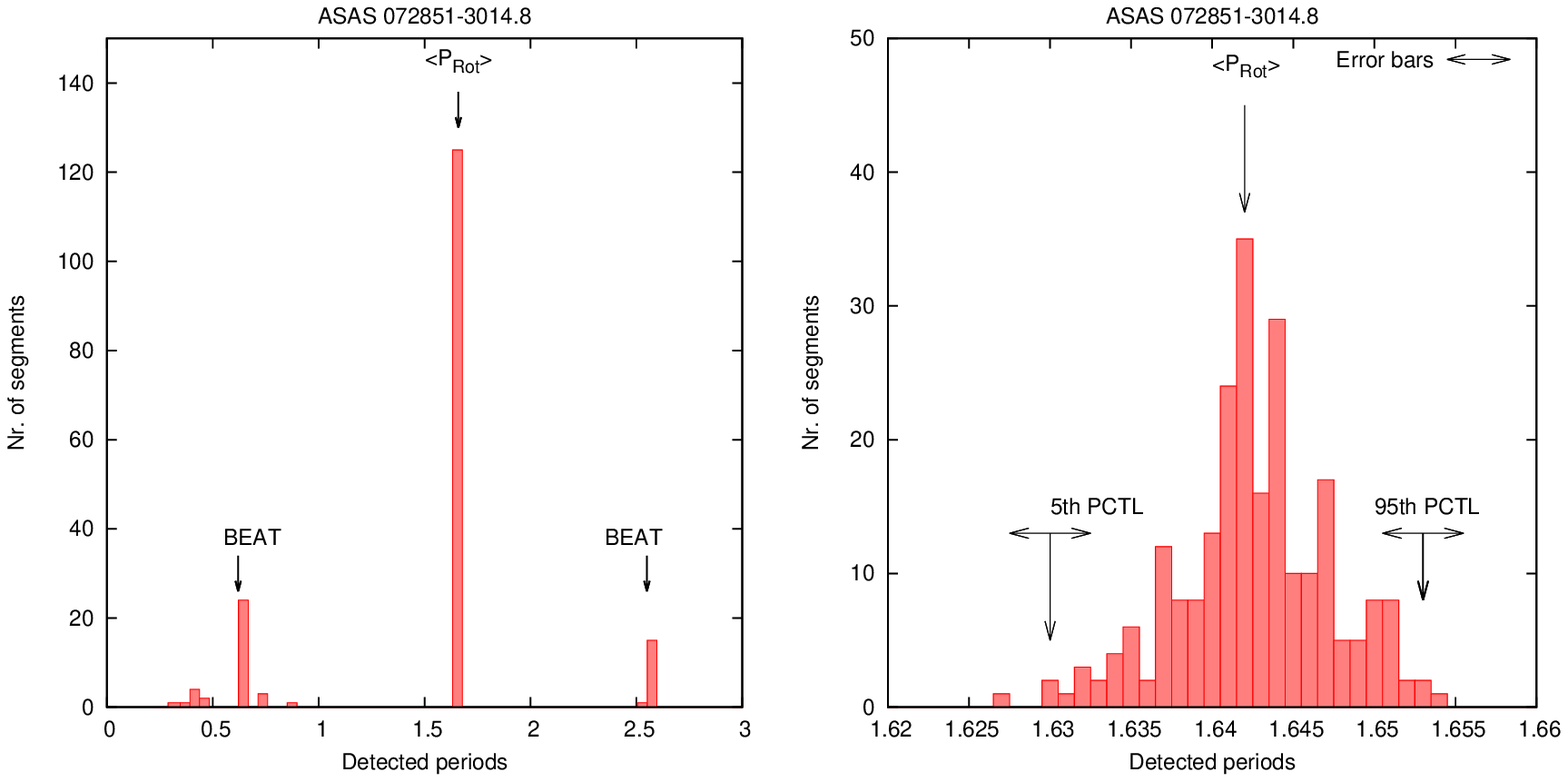}
\caption{The same of Fig. \ref{distribution_1} for the star ASAS J072851-3014.8}
\label{distribution2}
\end{center}
\end{figure*}

\begin{figure*}
\begin{center}
\includegraphics[]{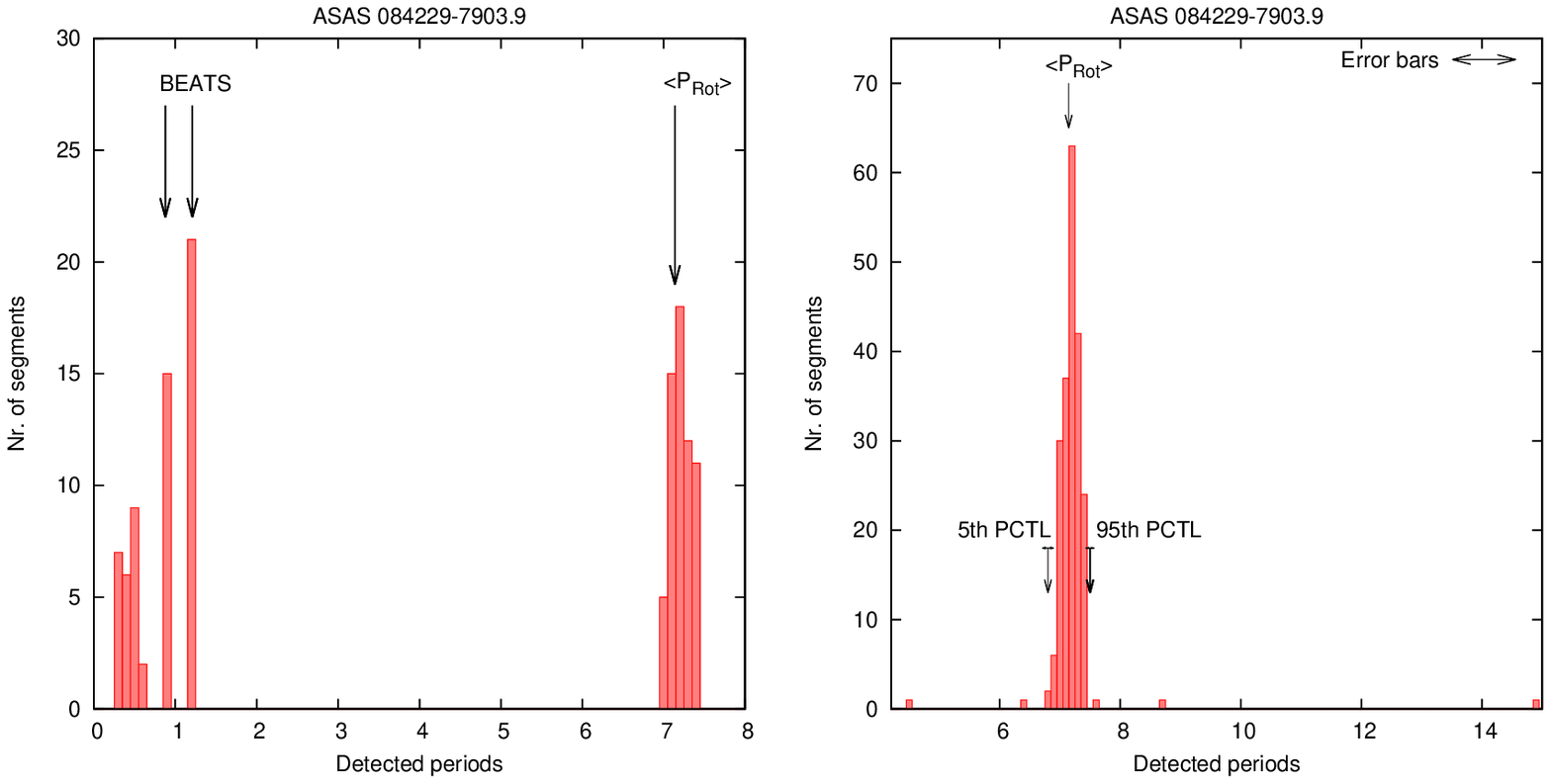}
\caption{The same of Fig. \ref{distribution_1} for the star ASAS J084229-7903.0}
\label{distribution3}
\end{center}
\end{figure*}

\begin{figure*}
\begin{center}
\includegraphics[]{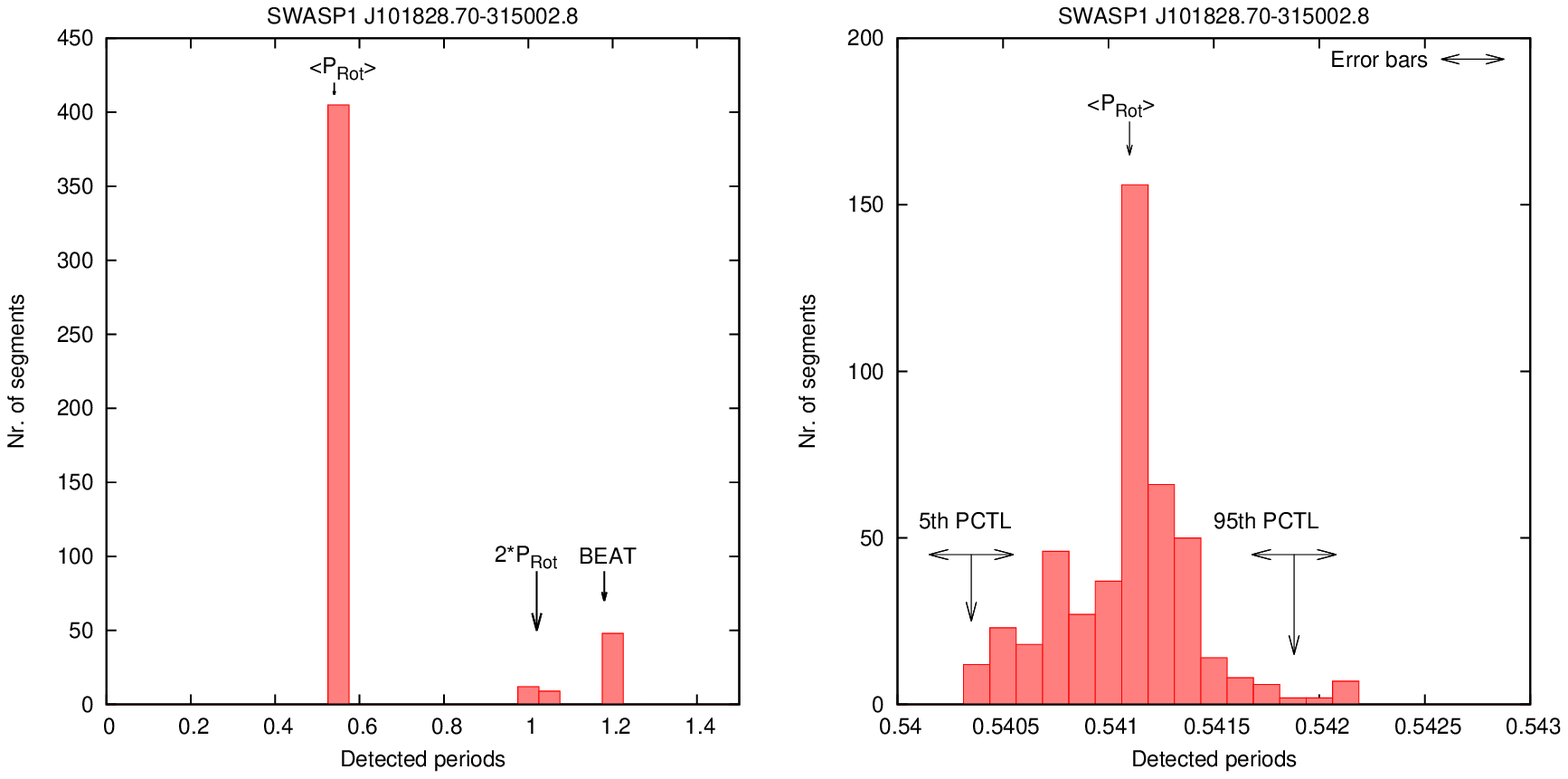}
\caption{The same of Fig. \ref{distribution_1} for the star SWASP1 J101828.70-315002.8 }
\label{distribution4}
\end{center}
\end{figure*}

\subsection{False Alarm Probability estimate}
We compute the False Alarm Probability  associated with a given period P by using the  \citet{1998MNRAS.301..831S} $\beta$ distribution:
\begin{equation}
\label{FAP}
Pr[Z_{\rm max}\le z] = 1 - (1 - (1 - 2z)^{(N - 3)/2} )^M
\end{equation}
where  Z is the power related to P in the periodogram, $N$ is the number of  data-points  and $M$ the number of independent frequencies, i.e. the number  of frequencies  at which the periodogram powers are independent random variables.
Unfortunately,  in the case of an unevenly sampled time-series, $M$ cannot be theoretically derived.
In this work we use the value estimated by  \citet{2012MNRAS.421.2774D}:
\begin{equation}
\label{M}
M~=1.8N_{\rm f}/k
\end{equation}
where $N_{\rm f}$ is the number of inspected frequencies and $k$ is a factor introduced to take into account the periodogram oversampling and defined as  $k=(1/T)/\delta\nu$ where T is the time interval spanned by the segment and $\delta\nu=0.0001 \rm d^{-1}$ the frequency step adopted to sample the periodogram.
Eq.  (\ref{M}) was derived  by \citet{2012MNRAS.421.2774D} by fitting empirical distributions of the peak powers $Z$ as generated by  Monte Carlo simulations.
In our analysis we use only periods with $FAP \le 0.01$. 

\subsection{Error estimate}
\label{errorcomputation}
The error associated with $\Delta\Omega_{\rm phot}$ is given by
\begin{equation}
\sigma^2_{\Delta\Omega_{\rm phot}} = \sigma^2_{\Omega_{\rm max}}+ \sigma^2_{\Omega_{\rm min}} 
\end{equation}
where $\sigma_{\Omega_{\rm max}}=2\pi\sigma_{\nu_{\rm max}}$ and $\sigma_{\Omega_{\rm min}}=2\pi\sigma_{\nu_{\rm min}}$.

The error associated with the frequency detected by the Lomb-Scargle periodogram has two components.
One component is due to the limited and discrete sampling of the signal whereas the other is due to the data noise.
\citet{1981Apkov81SS..78..175K} estimates the error $\delta\nu_{\rm samp}$ introduced by the sampling for a noise--free sinusoidal signal with frequency $\nu_0$ and  finds that this error is inversely correlated with $\nu_0T^2$ where $T$ is the interval spanned by the time-series.  \citet{1985PASP...97..285G} show that the result found by \citet{1981Apkov81SS..78..175K} can be approximated by:
\begin{equation}
\label{samplingerror}
\delta\nu_{\rm samp}=\frac{0.16}{\sqrt{2} \nu_0T^2}
\end{equation}
\citet{1981Apkov81SS..78..175K} derives this error for a signal uniformly sampled but the result can be applied also to unevenly sampled time-series. 
\citet{1981Apkov81SS..78..175K} makes also an estimate of the error  caused by Gaussian white noise as :
\begin{equation}
\label{gaussianerror}
\delta\nu_{\rm noise}=\frac{3\sigma}{4\sqrt{N}\nu_0TA}
\end{equation}
where $\sigma$ is the standard deviation of the noise, $N$ the number of time-series points and $A$ the amplitude of the signal.
Eq. (\ref{gaussianerror}) has been derived for a white Gaussian noise and can lead to an underestimation of the true error.
In order to make a more realistic estimate of the error due to the data noise we use the following approach:
\begin{itemize}
\item{we fit the data with a sinusoid; }
\item{we compute the fit residuals in order to have an estimate of the data noise;  }
\item{we make 10000 permutations of the residuals and we construct 10000 synthetic time-series by adding the permuted residual to the sinusoid }
\item{we run the period search algorithm on each synthetic time-series; }
\item{we build the distribution of the detected frequencies and take the standard deviation of this distribution as an estimate of the error.}
\end{itemize} 

The error  $\delta\nu_{\rm samp}$ increases with the stellar rotation period (see Eq. \ref{samplingerror}).  It is very small and ranges from $10^{-4}~\rm rad ~d^{-1}$, for stars rotating in 0.5 d,  to $0.006~ \rm rad ~ d^{-1}$ for stars rotating in 20 d (that is the highest rotation period detected in our target stars). 

The error  $\delta\nu_{\rm noise}$ inferred from the simulation of synthetic time-series depends on different factors like the stellar magnitude, the amplitude of variability and the number of observations.
It ranges between $0.003~ \rm rad ~d^{-1}$ and $0.02~ \rm rad ~d^{-1}$.

The final errors on $\Delta\Omega_{\rm phot}$ measurements range between $0.005$ and $0.025 ~ \rm rad~ d^{-1}$.

\subsection{Output of the analysis procedure}
In Figs. 6-9 we show the typical output of our analysis  for the  stars  ASAS J070030-7941.8, ASAS J072851-3014.8, ASAS J084229-7903.9 and SWASP1 J101828.70-315002.8

\begin{figure*}
\begin{center}
\includegraphics[angle=-90,width=140mm]{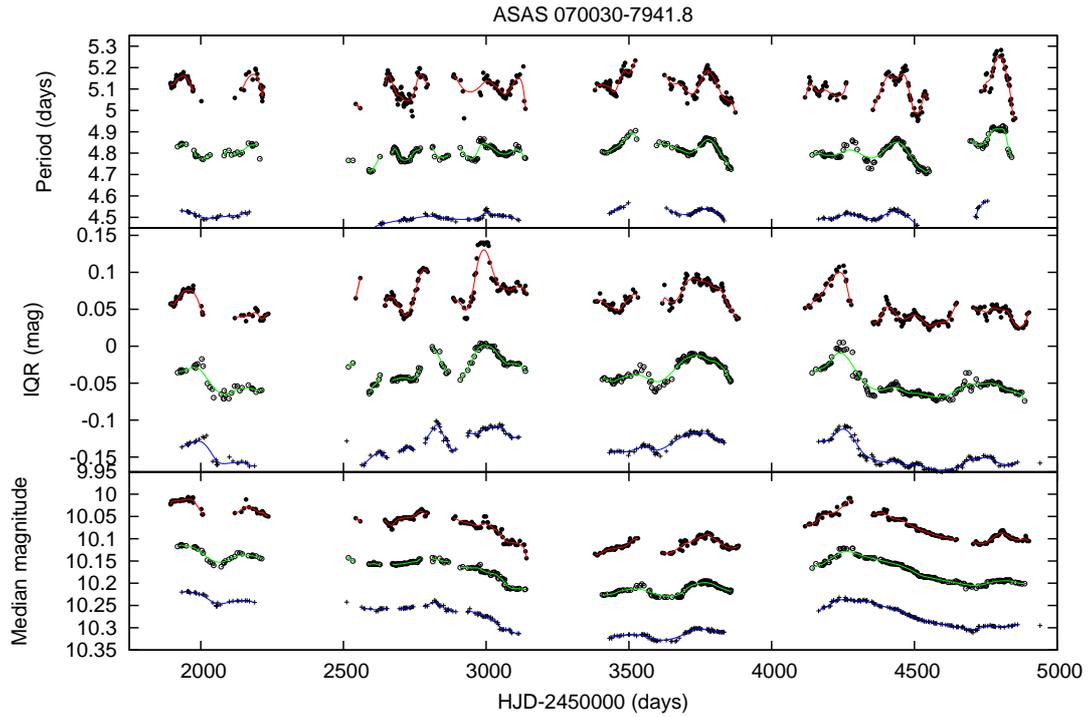}
\caption{The output of our analysis procedure in the case of the star with ASAS ID J070030-7941.8. The filled circles, the empty circles and the crosses are used to mark the results obtained with 50-d , 100-d and  150-d sliding windows, respectively. The results obtained with the 100-d and 150-d sliding window have been vertically shifted  to make the plot clearer.}  Top panel: rotation periods found in the different segments vs. time. Each point is located at the mid time of the corresponding segment. Mid panel: IQR measured in the different segments vs. time. Bottom panel: Median magnitude vs. time. The solid lines were obtained by fitting the data with smoothing cubic splines.
\label{output1}\end{center}
\end{figure*}

\begin{figure*}
\begin{center}
\includegraphics[angle=-90,width=140mm]{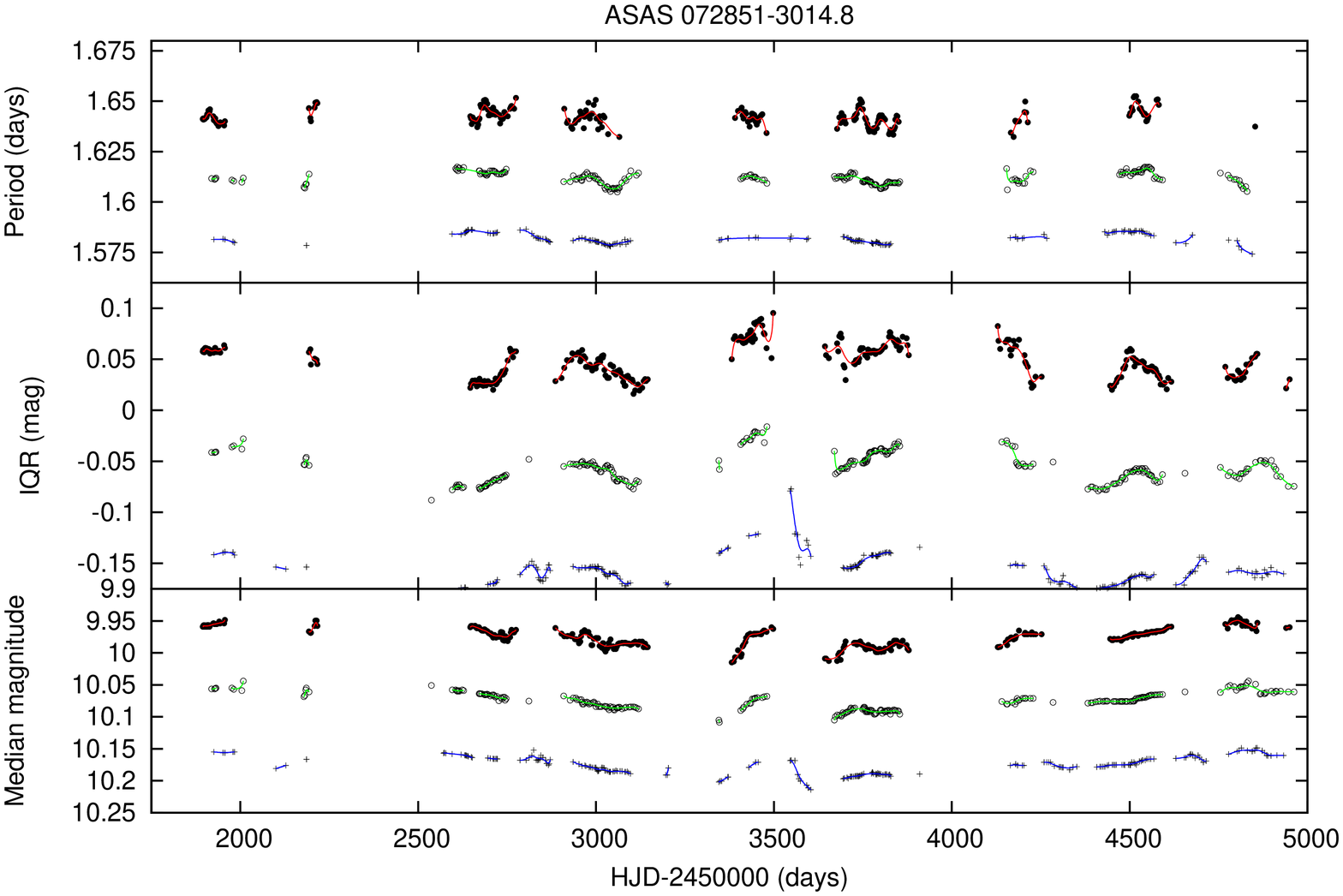}
\caption{The same of Fig. \ref{output1} for the star ASAS J072851-3014.8}
\label{output2}\end{center}
\end{figure*}

\begin{figure*}
\begin{center}
\includegraphics[angle=-90,width=140mm]{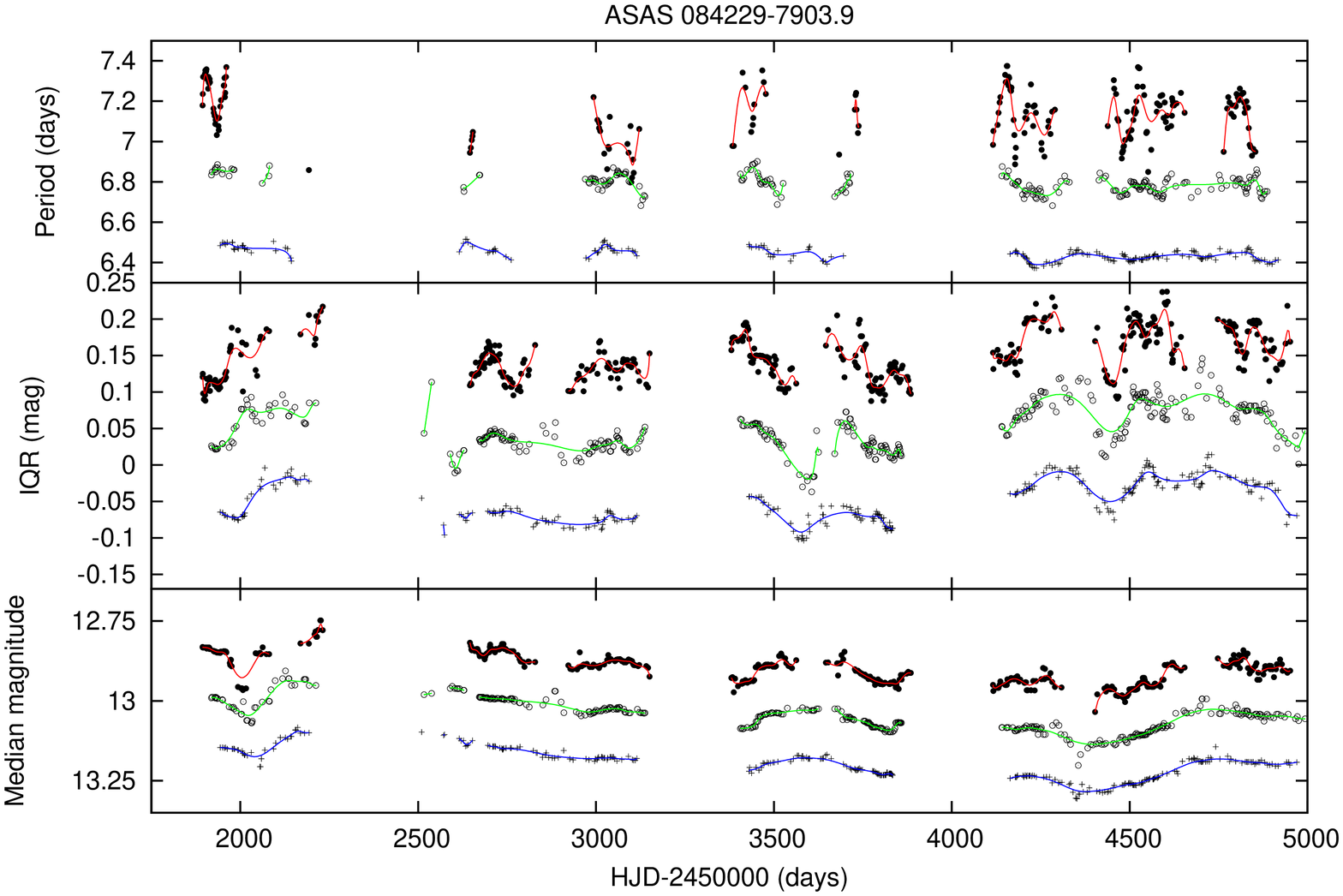}
\caption{The same of Fig. \ref{output1} for the star ASAS J0842.29-7903.9}
\label{output3}\end{center}
\end{figure*}

\begin{figure*}
\begin{center}
\includegraphics[angle=-90,width=140mm]{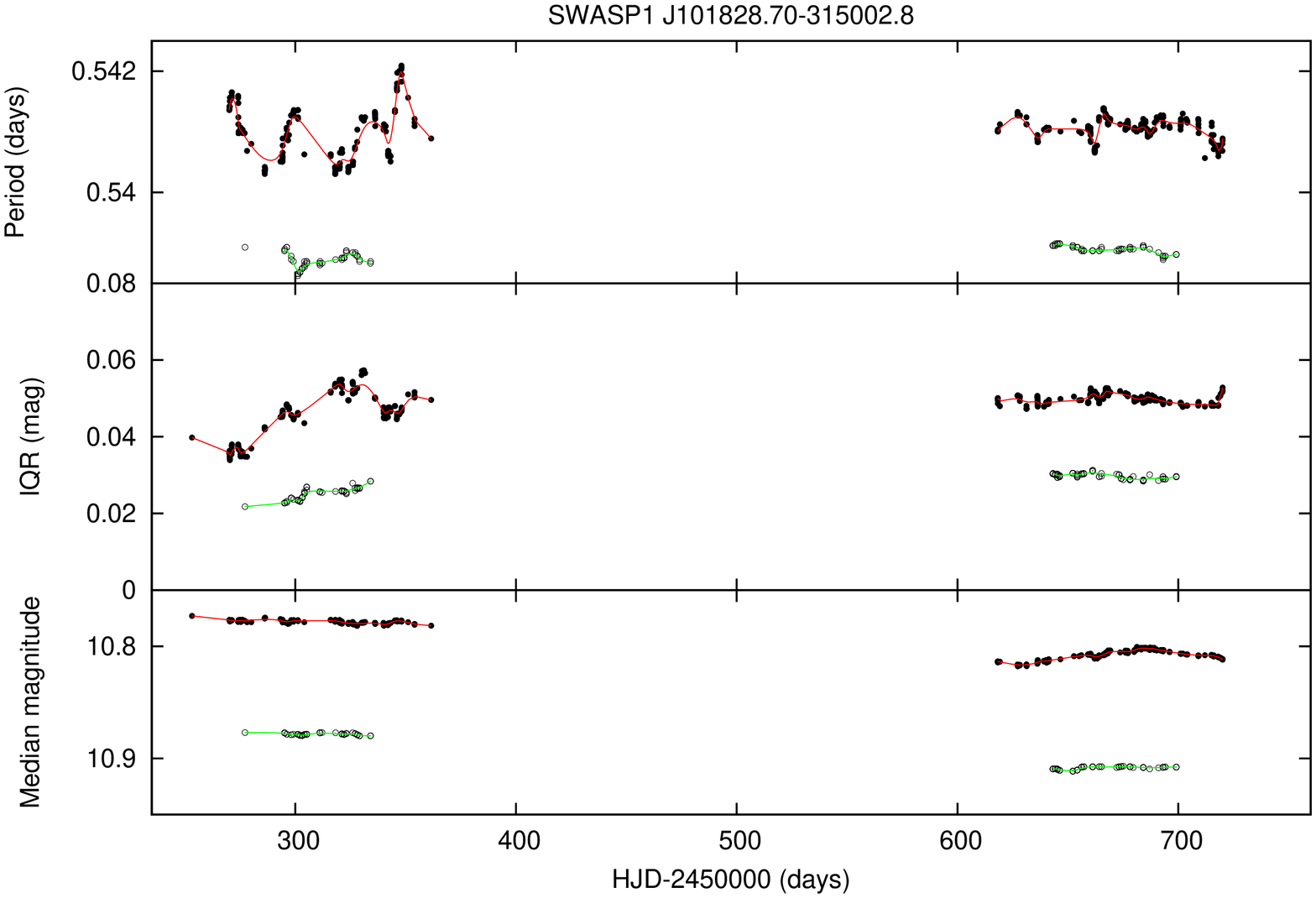}
\caption{The same of Fig. \ref{output1} for the star SWASP1 J101828.70-315002.8. The analysis with a 150-d sliding window has not been performed in this case because the two observation seasons span are not long enough.}
\label{output4}\end{center}
\end{figure*}

In the top panels  we plot the rotation periods detected in the different segments vs. the mid-observation times of the same segments. 
In the central panels we plot the Interquartile Range (IQR) of the observed magnitudes, that can be regarded as an index of the variability amplitude, and in the bottom panels we plotted  the median magnitude. 
The solid lines have been obtained by fitting the data with smoothing cubic splines and have been plotted in order to help the eye to follow the data trend.
The period and the IQR seem to follow well defined patterns that suggest "cyclical" changes in the ARs configurations. 
These cyclical patterns will be investigated in a forthcoming work.
In some segments,  no significant periods have been detected. This can be ascribed to different reasons:
\begin{itemize}
\item{the amplitude of the rotational modulation is  of the same order of the photometric error and therefore it is too low to be detected; }
\item{the ARs evolution occurs on a time scale comparable to or shorter than the length of the sliding window and therefore it masks the variability due to rotational modulation.}
\end{itemize}

In Figs. 6-9 we also plot the results obtained by using  100-d and  150-d  sliding windows for comparison. In principle, the use of the longer windows should give more precise measurements of the stellar rotation period as discussed in Sect. 3.4. In fact, the Equations shown  in Sect. 3.4 are strictly valid only for a pure sinusoidal signal. In the case of solar-like variables, the intrinsic evolution of ARs  tends to modify the amplitude and the phase of the  rotational signal and the use of  the longer windows does not improve the precision of the period measurements. Moreover, the use of a  given window tends to filter out the variability phenomena occurring in a time-scale comparable with or shorter than the window length. This produces smoother curves and flattens the amplitude of the  period variations. The use of a 50-d window is therefore more suitable to take into account the ARs evolution. Ideally, the length of the  window should cover a small number of stellar rotations. \citet{2015A&A...583A.134F},  for instance, suggest that long-term time-series of solar-like variables should be segmented in sub-series of length $5\times P_{Rot}$  in order to limit the effects of ARs evolution. Unfortunately, a similar choice of the window length is  unfeasible for the majority of our targets because of the sparse time-series sampling.

 \subsection{Comparison with other methods}
In recent years, several methods have been developed to measure SDR in solar-like stars.
Basically, these methods can be grouped in two classes.
The first class comprises methods based on Doppler Imaging \citep[see e.g.][]{1997MNRAS.291..658D,1997MNRAS.291....1D,2009A&ARv..17..251S,2011MNRAS.413.1949W} whereas the second comprises methods based on the analysis of photometric data \citep[see e.g.][]{1996ApJ...466..384D,2013A&A...560A...4R,2014A&14A...564A..50L}.
The former are very effective in measuring SDR in fast rotating stars ($v\rm sin \it i \ge \rm 15~ kms^{-1}$) but cannot be successfully applied to slow rotators.
The latter can be applied to slow rotators, but allow only the estimation of a lower limit for SDR.
In this section, we discuss the main differences between our  method and those employed by  \citet{2011MNRAS.413.1949W} and  \citet{2013A&A...560A...4R} taken as representative examples of the two classes.

\subsubsection{Comparison with Zeeman Doppler Imaging }
\citet{2011MNRAS.413.1949W} analyze  spectro-polarimetric and photometric data, acquired during nine nights, to study the topology of the magnetic field and the differential rotation of HD 106506. They complement the information inferred from the Zeeman Doppler Imaging (ZDI) with the photometric data and derive an equatorial rotation period $P_0=1.39\pm0.01~\rm d$ and a photospheric shear $\Delta\Omega=0.21\pm0.03~\rm rad~d^{-1}$.
They also study the spots distribution  on the stellar surface and find a large polar spot coupled with low and mid latitude features.   
 In order to compare our detection method with that employed by \citet{2011MNRAS.413.1949W}, we apply our analysis to the ASAS time-series of  $\rm HD ~106506$ 

The shortest detected period  is $P_{\rm min}=1.39~ \rm d$ that is equal to the equatorial period found by \citet{2011MNRAS.413.1949W}, while the  longest period is $P_{\rm max}=1.445~ \rm d$. 
The shear inferred from $P_{\rm min}$ and $P_{\rm max}$ is $\Delta\Omega_{\rm phot}=0.17~\rm{rad~ d ^{-1}}$ that is about 20 per cent less  than the value found by \citet{2011MNRAS.413.1949W}.
This difference is expected in view of the fact that  the photometric analysis fails to detect  the stellar rotation frequency at high latitudes because polar spots are  always visible and do not induce an appreciable flux modulation.

\subsubsection{Comparison with  short-term photometry analysis }
\citet{2013A&A...560A...4R}  measure a lower limit of SDR for about 20000 late--type stars by  analyzing the Q3 long--cadence  photometric time-series collected by the  Kepler mission.
These time-series cover an interval of about 90 days and therefore are not suitable for segmentation.
\citet{2013A&A...560A...4R} compute the Lomb-Scargle periodogram for the whole 90-d time-series and  estimate the amplitude of SDR  as 
\begin{equation}
\label{kepsdr}
 \Delta \Omega_{\rm phot} = \Omega_1 - \Omega_2
 \end{equation}
  where $\Omega_1=\frac{2\pi}{P_1}$, $\Omega_2=\frac{2\pi}{P_2}$ ,and  $P_1$ and $P_2$ are the two highest significant periods detected in the periodogram.
Eq. (\ref{kepsdr}) is based on the assumption that $P_1$ and $P_2$ are the rotation periods of two AR complexes located at different latitudes $\theta_1$ and $\theta_2$. 
This kind of measurement has two drawbacks:
\begin{itemize}
\item{the latitude range covered by $\theta_1$ and $\theta_2$ depends on the phase of the stellar magnetic cycle at which the time-series was collected. In fact, if a star has a solar-like cycle, the  ARs  gradually migrate from high latitudes (at the minimum of cycle)  to the equator (at the maximum);}
\item{it is based on the detection of multi periodicities in the same time-series and assumes that the secondary period $P_2$ is due to rotational modulation.  However , as \citet{2015MNRAS.450.3211A}  point out, the secondary peak in the Lomb-Scargle periodogram could be  induced also by an intrinsic evolution of the ARs.    }
\end{itemize}
The method employed here has two advantages over the technique used by \citet{2013A&A...560A...4R}.
First of all, it is based on the analysis of long-term photometry. The use of ASAS and Super-WASP time-series permits us to study our target stars over a time-scale comparable with the activity cycle duration and  to track stellar rotation in a wider  range of latitudes. This reduces the measurement bias due to the use of short-term photometry. 
The second advantage is that our method is not based on multiple periodicities, but searches for a drift in the primary period over the whole time-series. 
\citet{2015MNRAS.450.3211A} test different measurement methods on simulated light curves  and show that the use of multiple peaks can sometimes fail in detecting and measuring SDR.   The same authors point out that a more robust estimate of SDR can be done by searching for a  drift in the mean period over an activity cycle.

\section{Results}
In Table \ref{results} we report our $\Delta\Omega_{\rm phot}$ measurements. If for a given source both ASAS and SuperWASP estimates are available, the highest $\Delta\Omega_{\rm phot}$ value is reported. In the last column of the table we report a flag indicating whether the photometric shear has been computed with ASAS or SuperWASP data.

\subsection{Correlation between SDR and global stellar parameters}
We study how our $\Delta\Omega_{\rm phot}$ measurements are related to the  astrophysical parameters of our targets.
The astrophysical parameters are inferred by using the infrared magnitudes $M_J$ and $M_H$ and  the theoretical   isochrones developed by \citet{1998A&A...337..403B}, \citet{2000A&A...358..593S} and \citet{2013ApJ...776...87S}. 
We use infrared magnitudes because  the flux contrast between the spots  and photosphere is lower in these wavelengths passbands. Hence these magnitudes are less affected by the variability induced by rotational modulation in comparison with the optical magnitudes.  
For a given star, $M_J$ and $M_H$ are computed by adding the distance modulus $DM$ to  the   $J$ and $H$ magnitudes reported in the 2MASS catalogue. The distance modulus is computed by using the stellar parallax or the distance reported in the literature.
We do not apply any correction for reddening because the difference between the observed colors and those expected from the spectral type  is within the photometric errors of the 2MASS photometry.  
The expected colors have been taken by the list compiled by \citet{2013ApJS..208....9P} for the pre-main-sequence stars of different spectral types.   In Table \ref{input} we report the distance moduli and the inferred $M_J$ and $M_H$ magnitudes.
In Table \ref{parameters} we report, for each star, the mass and the effective temperature  inferred by the comparison with the different models. We report also the global convective turnover time-scale $\tau_{\rm C}$  inferred from the models of \citet{2013ApJ...776...87S}   and the derived Rossby Number $R=\frac{P_{\rm rot}}{\tau_{\rm C}}$.
The global convective turnover time-scale represents the characteristic time for the rise of a convective element through the stellar convection zone \citep[see appendix A of][for details on its computation]{2013ApJ...776...87S}.

\subsubsection{Correlation between SDR and  stellar temperature }

\begin{figure*}
\begin{center}
\includegraphics[angle=0]{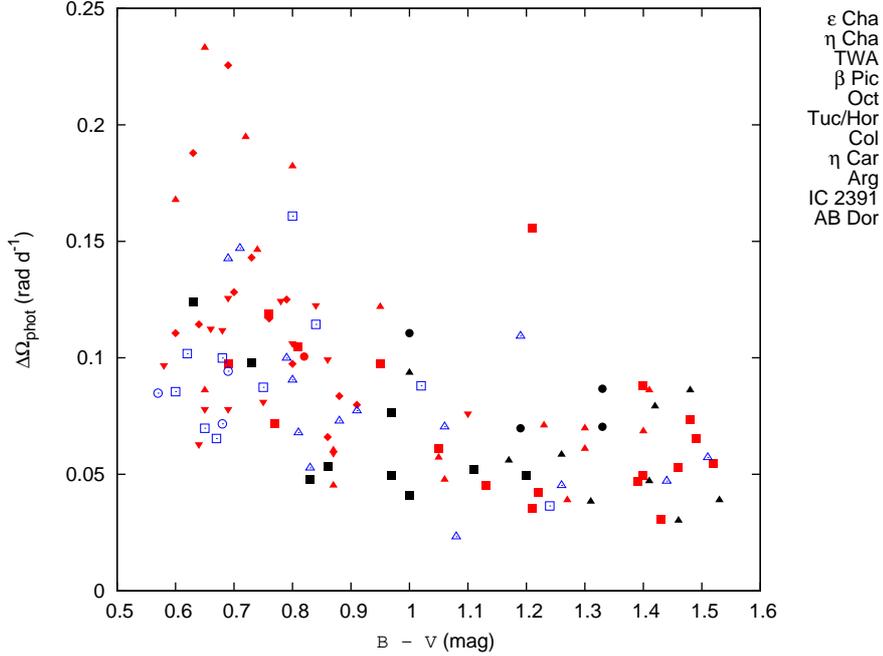}
\caption{The quantity $\Delta\Omega_{\rm phot}$ measured in the present work vs. the color index $B-V$. $\Delta\Omega_{\rm phot}$ increases toward bluest stars. The different symbols are used to mark stars belonging to different associations. }
\label{result}
\end{center}
\end{figure*}

In Fig. \ref{result} we plot $\Delta\Omega_{\rm phot}$ vs. the color index $B-V$ of our targets. Different symbols are used to mark stars belonging to different associations. Though the data are quite scattered, we note a general trend with SDR amplitude increasing toward bluest colors. 
The scatter can be ascribed to two different reasons. First of all, it is related to the intrinsic limitations of our measurement method  that allows only the detection of a lower limit for SDR (see discussion in Sect. 3). The second reason is that the plot mixes stars with different ages and rotation periods.   
Despite the scatter, the trend shown in the picture is in agreement with the theoretical models developed by \citet{1999A&A...344..911K} and \citet{2011AN....332..933K}.
In fact, in these models, the amplitude of stellar SDR is inversely correlated with the depth of the convective zone and, therefore, it increases toward higher temperatures.

In order to make a quantitative comparison between our results and those predicted by \citet{2011AN....332..933K}, we study the correlation between $\Delta\Omega_{\rm phot}$ and the stellar temperature.
 \begin{figure*}
\begin{center}
\includegraphics[scale=1]{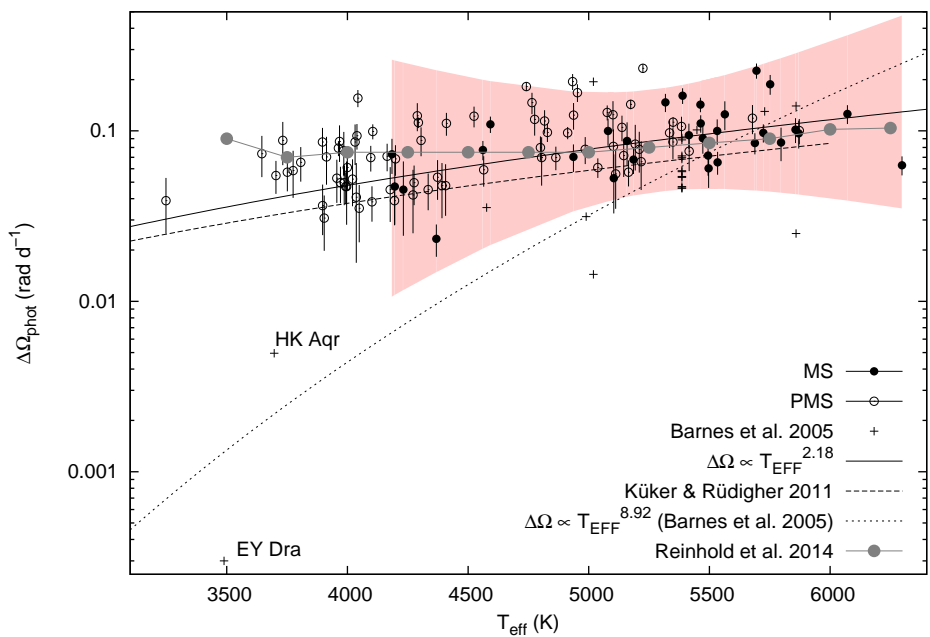}
\caption{ $\Delta\Omega_{\rm phot}$ vs. the effective temperature $T_{\rm eff}$. The black bullets mark the values measured in the present work.  The filled bullets mark stars that have reached ZAMS and the empty bullets mark stars that are still contracting. The black crosses mark the data reported in \protect\citet{2005MNRAS.357L...1B}. The continuous and the dotted line depict the power law inferred in the present work and that inferred in \protect\citet{2005MNRAS.357L...1B}, respectively. The dashed  line indicates the power law predicted by the theoretical model of \protect\citet{2011AN....332..933K}.  The $T_{\rm eff}$ values have been inferred from the  \protect\citet{2000A&A...358..593S} models. The shaded area represents the uncertainty in our power law fitting. Large gray bullets and grey line represents the results of \protect\cite{2013A&A...560A...4R}}
\label{dotemp}
\end{center}
\end{figure*}

\begin{longtab}
\begin{longtable}{lllllll}
\caption{\label{input}Data used to compute absolute magnitudes}\\
\hline\hline
 Target ID & parallax & distance & DM    & $M_J$  & $M_H$ & Ref.   \\
               &      ( mas) & ( pc )   & ( mag )  & ( mag )   &     ( mag )       &           \\ 
 \hline
\endfirsthead
\caption{continued.}\\
\hline\hline
 Target ID & distance & parallax & DM    & $M_J$  & $M_H$ & Ref.   \\
               &      ( pc ) & ( mas )   & ( mag )  & ( mag )   &     ( mag )       &           \\ 
 \hline
\endhead
\hline
\endfoot
    ASAS J001353-7441.3 & 22.8 & 43.86 & 3.21 & 4.20 & 3.88 & 1\\
  ASAS J002409-6211.1 & 21.6 & 46.30 & 3.33 & 5.06 & 4.38 & 1\\
  ASAS J003451-6155.0 & 21.8 & 45.87 & 3.31 & 4.03 & 3.41 & 1\\
  ASAS J004220-7747.7 &   & 50.00 & 3.49 & 4.73 & 4.16 &  2\\
  ASAS J011315-6411.6 & 17.1 & 58.48 & 3.83 & 4.78 & 4.29 & 3\\
  ASAS J015749-2154.1 & 23.6 & 42.37 & 3.14 & 3.72 & 3.42 & 1\\
  ASAS J020136-1610.0 &   & 92.00 & 4.82 & 3.79 & 3.28 &  2\\
  ASAS J020718-5311.9 & 19.9 & 50.25 & 3.51 & 3.84 & 3.48 & 1\\
  ASAS J024126+0559.3 & 24.7 & 40.49 & 3.04 & 4.87 & 4.20 & 1\\
  ASAS J024233-5739.6 &   & 50.00 & 3.49 & 5.07 & 4.47 &  2\\
  ASAS J030942-0934.8 & 26.73 & 37.41 & 2.87 & 4.29 & 3.93 & 3\\
  ASAS J031909-3507.0 &   & 44.00 & 3.22 & 5.36 & 4.70 &  2\\
  ASAS J033049-4555.9 &   & 44.00 & 3.22 & 4.55 & 4.03 &  2\\
  ASAS J033156-4359.2 &   & 42.00 & 3.12 & 5.18 & 4.56 &  2\\
  ASAS J034723-0158.3 & 62.0 & 16.13 & 1.04 & 6.77 & 6.14 & 3\\
  ASAS J045249-1955.0 &   & 72.00 & 4.29 & 3.76 & 3.21 &  2\\
  ASAS J045305-4844.6 &   & 76.00 & 4.40 & 4.46 & 4.02 &  2\\
  ASAS J045935+0147.0 & 37.5 & 26.67 & 2.13 & 4.99 & 4.32 & 1\\
  ASAS J050047-5715.4 & 38.1 & 26.25 & 2.10 & 5.00 & 4.33 & 1\\
  ASAS J050230-3959.2 &   & 42.00 & 3.12 & 5.61 & 5.09 &  2\\
  ASAS J050651+7221.2 &   & 143.00 & 5.78 &   &   &  2\\
  ASAS J052845-6526.9 & 66.9 & 14.95 & .87 & 4.44 & 3.97 & 1\\
  ASAS J052857-3328.3 &   & 57.00 & 3.78 & 4.65 & 4.14 &  2\\
  ASAS J053705-3932.4 & 18.87 & 52.99 & 3.62 & 4.28 & 3.84 & 3\\
  ASAS J055101-5238.2 &   & 106.00 & 5.13 & 3.96 & 3.56 &  2\\
  ASAS J055329-8156.9 &   & 59.00 & 3.85 & 3.68 & 3.19 &  2\\
  ASAS J055751-3804.1 &   & 68.00 & 4.16 & 3.97 & 3.58 &  2\\
  ASAS J060834-3402.9 &   & 72.00 & 4.29 & 4.43 & 4.04 &  2\\
  ASAS J061828-7202.7 & 25.9 & 38.61 & 2.93 & 4.60 & 4.05 & 1\\
  ASAS J062607-4102.9 &   & 91.00 & 4.80 & 3.74 & 3.36 &  2\\
  ASAS J062806-4826.9 &   & 135.00 & 5.65 & 3.88 & 3.46 &  2\\
  ASAS J063950-6128.7 & 45.6 & 21.93 & 1.71 & 5.60 & 4.94 & 1\\
  ASAS J064346-7158.6 & 17.7 & 56.50 & 3.76 & 3.93 & 3.62 & 1\\
  ASAS J065623-4646.9 &   & 75.00 & 4.38 & 4.01 & 3.52 &  2\\
  ASAS J070030-7941.8 & 15.6 & 64.10 & 4.03 & 4.23 & 3.80 & 1\\
  ASAS J070153-4227.9 & 10.7 & 93.46 & 4.85 &   &   & 5\\
  ASAS J072124-5720.6 &   & 100.00 & 5.00 & 4.21 & 3.78 &  2\\
  ASAS J072822-4908.6 & 12.6 & 79.37 & 4.50 & 4.01 & 3.63 & 5\\
  ASAS J072851-3014.8 & 64.2 & 15.58 & .96 & 5.65 & 5.01 & 1\\
  ASAS J073547-3212.2 & 28.29 & 35.35 & 2.74 & 4.16 & 3.84 & 3\\
  ASAS J082406-6334.1 &   & 104.00 & 5.09 & 3.48 & 3.15 &  2\\
  ASAS J082844-5205.7 &   & 120.00 & 5.40 & 3.92 & 3.66 &  2\\
  ASAS J083656-7856.8 &   & 99.00 & 4.98 & 3.18 & 2.52 &  2\\
  ASAS J084006-5338.1 &   & 139.00 & 5.72 & 3.61 & 3.32 &  2\\
  ASAS J084200-6218.4 &   & 147.00 & 5.84 & 3.56 & 3.13 & 2\\
  ASAS J084229-7903.9 &   & 97.30 & 4.94 & 4.59 & 3.84 & 4\\
  ASAS J084300-5354.1 &   & 149.00 & 5.87 & 3.89 & 3.51 &  2\\
  ASAS J084432-7846.6 &   & 97.30 & 4.94 & 4.71 & 3.98 & 4\\
  ASAS J084708-7859.6 &   & 97.30 & 4.94 & 3.79 & 3.09 & 4\\
  ASAS J085005-7554.6 &   & 101.00 & 5.02 & 4.24 & 3.83 &  2\\
  ASAS J085156-5355.9 &   & 141.00 & 5.75 & 3.57 & 3.15 &  2\\
  ASAS J085746-5408.6 &   & 159.00 & 6.01 & 3.90 & 3.47 &  2\\
  ASAS J085752-4941.8 &   & 111.00 & 5.23 & 3.86 & 3.53 & 2\\
  ASAS J085929-5446.8 &   & 106.00 & 5.13 & 3.66 & 3.33 &  2\\
  ASAS J092335-6111.6 & 11.8 & 84.75 & 4.64 & 3.92 & 3.42 & 1\\
  ASAS J092854-4101.3 & 6.7 & 149.25 & 5.87 & 3.92 & 3.40 & 5\\
  ASAS J094247-7239.8 & 6.5 & 153.85 & 5.94 & 4.51 & 4.07 & 5\\
  ASAS J095558-6721.4 & 9.2 & 108.70 & 5.18 & 3.52 & 3.20 & 5\\
  ASAS J101315-5230.9 &   & 48.00 & 3.41 & 4.46 & 3.95 &  2\\
  ASAS J105351-7002.3 & 6.3 & 158.73 & 6.00 & 3.43 & 3.15 & 5\\
  ASAS J105749-6914.0 &   & 112.00 & 5.25 & 3.24 & 2.76 & 6\\
  ASAS J110914-3001.7 &   & 41.00 & 3.06 & 4.57 & 3.86 &  2\\
  ASAS J112105-3845.3 & 31.25 & 32.00 & 2.53 & 6.47 & 5.81 & 3\\
  ASAS J112117-3446.8 &   & 55.00 & 3.70 & 4.73 & 4.03 &  2\\
  ASAS J112205-2446.7 & 21.4 & 46.73 & 3.35 & 3.05 & 2.41 & 1\\
  ASAS J115942-7601.4 & 10.8 & 107.00 & 5.15 & 3.99 & 3.32 & 6\\
  ASAS J120139-7859.3 &   & 102.00 & 5.04 & 2.22 & 1.92 & 6\\
  ASAS J120204-7853.1 &   & 110.00 & 5.21 & 4.01 & 3.25 & 6\\
  ASAS J121138-7110.6 &   & 112.00 & 5.25 & 2.43 & 2.06 & 6\\
  ASAS J121531-3948.7 &   & 51.00 & 3.54 & 4.63 & 3.97 &  2\\
  ASAS J122023-7407.7 &   & 110.00 & 5.21 & 4.05 & 3.40 & 6\\
  ASAS J122034-7539.5 & 19.5 & 51.28 & 3.55 & 5.03 & 4.54 & 5\\
  ASAS J122105-7116.9 &   & 100.00 & 5.00 & 4.09 & 3.42 & 6\\
  ASAS J123921-7502.7 &   & 100.00 & 5.00 & 3.43 & 2.95 & 6\\
  ASAS J125826-7028.8 &   & 99.00 & 4.98 & 3.21 & 2.72 & 6\\
  ASAS J134913-7549.8 & 12.3 & 81.30 & 4.55 & 3.82 & 3.48 & 5\\
  ASAS J153857-5742.5 & 25.2 & 39.68 & 2.99 & 3.39 & 3.00 & 1\\
  ASAS J171726-6657.1 & 31.8 & 31.45 & 2.49 & 2.80 & 2.42 & 1\\
  ASAS J181411-3247.5 &   & 73.00 & 4.32 & 3.75 & 3.12 &  2\\
  ASAS J181952-2916.5 & 13.3 & 75.19 & 4.38 & 3.15 & 2.82 & 1\\
  ASAS J184653-6210.6 &   & 54.00 & 3.66 & 5.08 & 4.39 &  2\\
  ASAS J185306-5010.8 & 20.1 & 49.75 & 3.48 & 3.37 & 3.00 & 1\\
  ASAS J200724-5147.5 & 33.6 & 29.76 & 2.37 & 5.79 & 5.20 & 5\\
  ASAS J204510-3120.4 & 100.6 & 9.94 & -.01 & 5.45 & 4.84 & 1\\
  ASAS J205603-1710.9 &   & 47.00 & 3.36 & 4.49 & 3.89 &  2\\
  ASAS J212050-5302.0 & 21.8 & 45.87 & 3.31 & 4.08 & 3.72 & 1\\
  ASAS J214430-6058.6 & 23.6 & 42.37 & 3.14 & 5.62 & 4.95 & 1\\
  ASAS J232749-8613.3 &   & 60.00 & 3.89 & 4.06 & 3.70 &  2\\
  ASAS J233231-1215.9 &   & 28.00 & 2.24 & 5.21 & 4.53 &  2\\
  ASAS J234154-3558.7 & 16.0 & 62.50 & 3.98 & 4.12 & 3.79 & 1\\
  SWASP1 J002334.66+201428.6 &   & 31.00 & 2.46 & 5.68 & 5.04 &  2\\
  SWASP1 J021055.38-460358.6 &   & 70.00 & 4.23 & 5.06 & 4.53 &  2\\
  SWASP1 J022729.25+305824.6 & 25.03 & 39.95 & 3.01 & 4.86 & 4.23 & 3\\
  SWASP1 J033120.80-303058.7 & 7.8 & 128.21 & 5.54 & 2.93 & 2.58 & 1\\
  SWASP1 J041422.57-381901.5 & 12.1 & 82.64 & 4.59 & 3.36 & 3.11 & 3\\
  SWASP1 J042148.68-431732.5 &   & 143.00 & 5.78 & 3.14 & 2.76 &  2\\
  SWASP1 J043450.78-354721.2 &   & 77.00 & 4.43 & 4.86 & 4.32 &  2\\
  SWASP1 J045153.54-464713.3 &   & 80.00 & 4.52 & 4.03 & 3.68 &  2\\
  SWASP1 J050649.47-213503.7 &   & 18.00 & 1.28 & 5.77 & 5.11 &  2\\
  SWASP1 J051829.04-300132.0 &   & 65.00 & 4.06 & 5.07 & 4.38 &  2\\
  SWASP1 J052855.09-453458.3 &   & 78.00 & 4.46 & 5.10 & 4.61 &  2\\
  SWASP1 J053504.11-341751.9 &   & 78.00 & 4.46 & 5.34 & 4.79 &  2\\
  SWASP1 J054516.24-383649.1 &   & 90.00 & 4.77 & 4.79 & 4.38 &  2\\
  SWASP1 J055021.43-291520.7 &   & 149.00 & 5.87 & 3.77 & 3.38 &  2\\
  SWASP1 J064118.50-382036.1 &   & 78.00 & 4.46 & 5.02 & 4.48 &  2\\
  SWASP1 J101828.70-315002.8 & 12.99 & 76.98 & 4.43 & 4.44 & 3.75 & 3\\
  SWASP1 J113241.23-265200.7 &   & 39.00 & 2.96 &   &   &  2\\
  SWASP1 J114824.21-372849.2 & 19.9 & 50.25 & 3.51 & 5.18 & 4.53 & 1\\
  SWASP1 J191144.66-260408.5 &   & 80.00 & 4.52 & 3.57 & 3.04 &  2\\
  SWASP1 J224457.83-331500.6 & 42.4 & 23.58 & 1.86 & 5.92 & 5.29 & 1\\
  SWASP1 J231152.05-450810.6 & 19.7 & 50.76 & 3.53 & 3.94 & 3.58 & 1\\
 \end{longtable}
\tablefoot{For each target we reported the distance or the parallax reported in previous works, the distance modulus and the absolute J  H magnitudes inferred in the present work. In the last column we reported the reference from which the distance or the parallax has been taken. References: (1) \protect\citet{2006A&A...460..695T}; (2) \protect\citet{2008hsf2.book..757T}; (3) \protect\citet{1997A&A...323L..49P}; (4) \protect\citet{2000ApJ...544..356M}; (5) \protect\citet{2013MNRAS.431.1005D}; (6) \protect\citet{2013MNRAS.435.1325M}.}
\end{longtab}

\begin{longtab}
\begin{longtable}{lllllllll}
\caption{\label{parameters}Stellar parameters inferred by the comparison of J and H magnitudes with different theoretical models.   }\\
\hline\hline
Target ID &  $M_{\rm Siess}$ & $T_{\rm Siess}$ & $M_{\rm Baraffe}$ & $T_{\rm Baraffe}$  & $M_{\rm Spada}$ & $T_{\rm Spada}$ & $\tau_{\rm C}$& Rossby nr.   \\
               &   ( $\rm M_{\sun}$ ) &( K ) & ( \rm $M_{\sun}$ )  &( K )    &    ( $\rm M_{\sun}$ )        & ( K )  & ( d ) &          \\  
\hline
\endfirsthead
\caption{continued.}\\
\hline\hline
Target ID &  $M_{\rm Siess}$ & $T_{\rm Siess}$ & $M_{\rm Baraffe}$ & $T_{\rm Baraffe}$  & $M_{\rm Spada}$ & $T_{\rm Spada}$ & $\tau_{\rm C}$& Rossby nr.   \\
               &   ( $\rm M_{\sun}$ ) &( K ) & ( $\rm M_{\sun}$ )  &( K )    &    ( $ \rm M_{\sun}$ )        & ( K )  & ( d ) &          \\  
\hline
\endhead
\hline
\endfoot
  ASAS J001353-7441.3 & 0.92 & 4764.88 & 0.93 & 4464.05 & 0.95 & 5060.67 & 49.717 & 0.074\\
  ASAS J002409-6211.1 & 0.79 & 4199.02 & 0.8 & 3997.71 & 0.78 & 4325.44 & 146.045 & 0.012\\
  ASAS J003451-6155.0 & 0.99 & 5164.99 & 0.98 & 4727.36 & 1.03 & 5446.5 & 31.85 & 0.012\\
  ASAS J004220-7747.7 & 0.84 & 4407.14 & 0.86 & 4165.18 & 0.86 & 4608.72 & 92.809 & 0.028\\
  ASAS J011315-6411.6 & 0.83 & 4334.69 & 0.84 & 4109.04 & 0.83 & 4513.54 & 118.681 & 0.011\\
  ASAS J015749-2154.1 & 1.03 & 5348.77 & 1.0 & 4861.71 & 1.06 & 5596.57 & 24.489 & 0.124\\
  ASAS J020136-1610.0 & 1.05 & 5416.15 & 1.01 & 4897.35 & 1.07 & 5649.15 & 24.489 & 0.131\\
  ASAS J020718-5311.9 & 1.0 & 5224.06 & 0.99 & 4774.86 & 1.04 & 5499.37 & 31.85 & 0.073\\
  ASAS J024126+0559.3 & 0.78 & 4049.27 & 0.78 & 3827.04 & 0.83 & 4115.02 & 195.566 & 0.025\\
  ASAS J024233-5739.6 & 0.77 & 4164.55 & 0.79 & 3971.28 & 0.77 & 4263.07 & 146.045 & 0.051\\
  ASAS J030942-0934.8 & 0.96 & 5185.52 & 0.91 & 4982.03 & 0.95 & 5545.63 & 38.991 & 0.14\\
  ASAS J031909-3507.0 & 0.69 & 3999.56 & 0.73 & 3852.94 & 0.71 & 4000.66 & 171.965 & 0.05\\
  ASAS J033049-4555.9 & 0.87 & 4524.9 & 0.88 & 4272.77 & 0.89 & 4764.23 & 92.809 & 0.041\\
  ASAS J033156-4359.2 & 0.74 & 4097.0 & 0.77 & 3919.74 & 0.74 & 4151.49 & 171.965 & 0.017\\
  ASAS J034723-0158.3 & 0.45 & 3751.22 & 0.48 & 3627.29 & 0.49 & 3684.28 & 173.081 & 0.022\\
  ASAS J045249-1955.0 & 1.07 & 5496.24 & 1.02 & 4956.87 & 1.08 & 5709.79 & 24.489 & 0.213\\
  ASAS J045305-4844.6 & 0.88 & 4564.63 & 0.89 & 4311.88 & 0.9 & 4820.09 & 70.606 & 0.065\\
  ASAS J045935+0147.0 & 0.73 & 3992.85 & 0.74 & 3783.13 & 0.8 & 4015.72 & 195.566 & 0.023\\
  ASAS J050047-5715.4 & 0.73 & 3986.81 & 0.74 & 3778.59 & 0.8 & 4005.8 & 195.566 & 0.045\\
  ASAS J050230-3959.2 & 0.67 & 4184.49 & 0.64 & 4111.98 & 0.68 & 4320.15 & 75.111 & 0.088\\
  ASAS J050651-7221.2 & 1.2 & 5872.15 & 1.06 & 5236.3 & 1.13 & 5996.48 & 21.57 & 0.011\\
  ASAS J052845-6526.9 & 0.93 & 5103.88 & 0.88 & 4894.7 & 0.93 & 5455.8 & 44.701 & 0.011\\
  ASAS J052857-3328.3 & 0.88 & 4936.3 & 0.84 & 4752.25 & 0.88 & 5274.9 & 49.528 & 0.014\\
  ASAS J053705-3932.4 & 0.92 & 4741.35 & 0.92 & 4444.53 & 0.95 & 5038.04 & 49.717 & 0.05\\
  ASAS J055101-5238.2 & 0.98 & 5101.79 & 0.97 & 4689.95 & 1.02 & 5392.94 & 31.85 & 0.038\\
  ASAS J055329-8156.9 & 1.08 & 5564.02 & 1.02 & 5007.28 & 1.09 & 5760.17 & 24.489 & 0.076\\
  ASAS J055751-3804.1 & 1.04 & 5462.61 & 1.01 & 5292.01 & 1.04 & 5862.33 & 32.162 & 0.024\\
  ASAS J060834-3402.9 & 0.93 & 5079.95 & 0.88 & 4872.36 & 0.92 & 5427.25 & 44.701 & 0.076\\
  ASAS J061828-7202.7 & 0.87 & 4177.56 & 0.86 & 3918.0 & 0.88 & 4302.06 & 173.338 & 0.015\\
  ASAS J062607-4102.9 & 1.04 & 5384.07 & 1.0 & 4878.44 & 1.07 & 5623.8 & 24.489 & 0.171\\
  ASAS J062806-4826.9 & 1.0 & 5209.3 & 0.99 & 4766.8 & 1.04 & 5490.63 & 31.85 & 0.041\\
  ASAS J063950-6128.7 & 0.69 & 4231.9 & 0.66 & 4162.5 & 0.69 & 4383.69 & 75.111 & 0.121\\
  ASAS J064346-7158.6 & 0.97 & 5076.36 & 0.97 & 4674.16 & 1.02 & 5370.69 & 31.85 & 0.122\\
  ASAS J065623-4646.9 & 0.98 & 5100.6 & 0.97 & 4686.36 & 1.02 & 5390.86 & 31.85 & 0.14\\
  ASAS J070030-7941.8 & 0.93 & 4799.47 & 0.93 & 4482.09 & 0.96 & 5097.57 & 49.717 & 0.103\\
  ASAS J072124-5720.6 & 0.93 & 4817.17 & 0.93 & 4493.68 & 0.97 & 5115.58 & 49.717 & 0.094\\
  ASAS J072822-4908.6 & 1.02 & 5388.54 & 0.92 & 4920.61 & 1.02 & 5578.2 & 31.956 & 0.032\\
  ASAS J072851-3014.8 & 0.67 & 4196.56 & 0.65 & 4122.64 & 0.68 & 4336.36 & 75.111 & 0.022\\
  ASAS J073547-3212.2 & 0.94 & 5158.73 & 0.89 & 4724.24 & 0.97 & 5390.66 & 35.458 & 0.143\\
  ASAS J082406-6334.1 & 1.15 & 5751.56 & 1.04 & 5131.28 & 1.11 & 5886.94 & 21.575 & 0.037\\
  ASAS J082844-5205.7 & 1.03 & 5415.05 & 0.93 & 4972.31 & 1.03 & 5615.66 & 32.309 & 0.047\\
  ASAS J083656-7856.8 & 1.41 & 4863.64 & 1.36 & 4459.57 &  &  &  & \\
  ASAS J084006-5338.1 & 1.12 & 5688.16 & 1.09 & 5497.24 & 1.1 & 5945.58 & 21.027 & 0.064\\
  ASAS J084200-6218.4 & 1.14 & 5723.08 & 1.04 & 5103.57 & 1.11 & 5858.1 & 21.575 & 0.057\\
  ASAS J084229-7903.9 & 0.72 & 3967.34 & 0.67 & 3683.61 & 0.82 & 3974.45 & 306.563 & 0.023\\
  ASAS J084300-5354.1 & 1.05 & 5494.3 & 0.95 & 5104.5 & 1.05 & 5709.71 & 27.17 & 0.116\\
  ASAS J084432-7846.6 & 0.66 & 3913.33 & 0.62 & 3645.25 & 0.75 & 3870.19 & 338.808 & 0.06\\
  ASAS J084708-7859.6 & 1.14 & 4410.09 & 1.07 & 4063.22 & 1.12 & 4690.61 & 175.865 & 0.028\\
  ASAS J085005-7554.6 & 0.92 & 4776.04 & 0.93 & 4467.9 & 0.96 & 5073.41 & 49.717 & 0.023\\
  ASAS J085156-5355.9 & 1.13 & 5694.86 & 1.04 & 5089.11 & 1.11 & 5843.03 & 21.575 & 0.089\\
  ASAS J085746-5408.6 & 0.99 & 5192.59 & 0.98 & 4754.38 & 1.04 & 5475.28 & 31.85 & 0.061\\
  ASAS J085752-4941.8 & 0.99 & 5173.56 & 0.98 & 4741.8 & 1.03 & 5456.74 & 31.85 & 0.064\\
  ASAS J085929-5446.8 & 1.06 & 5463.98 & 1.01 & 4933.05 & 1.08 & 5683.11 & 24.489 & 0.018\\
  ASAS J092335-6111.6 & 1.0 & 5216.98 & 0.99 & 4769.42 & 1.04 & 5496.23 & 31.85 & 0.122\\
  ASAS J092854-4101.3 & 1.07 & 5533.54 & 1.0 & 5242.54 & 1.06 & 5751.23 & 27.07 & 0.014\\
  ASAS J094247-7239.8 & 0.86 & 4804.63 & 0.85 & 4479.31 & 0.9 & 5090.13 & 41.081 & 0.056\\
  ASAS J095558-6721.4 & 1.17 & 5796.62 & 1.12 & 5577.32 & 1.13 & 6066.08 & 21.867 & 0.083\\
  ASAS J101315-5230.9 & 0.79 & 4039.73 & 0.78 & 3778.3 & 0.9 & 4147.98 & 219.272 & 0.02\\
  ASAS J105351-7002.3 & 1.19 & 5856.48 & 1.14 & 5622.04 & 1.14 & 6140.48 & 21.867 & 0.047\\
  ASAS J105749-6914.0 & 1.16 & 4373.67 & 1.12 & 4012.34 & 1.2 & 4685.66 & 233.527 & 0.015\\
  ASAS J110914-3001.7 & 0.79 & 4031.79 & 0.77 & 3774.52 & 0.9 & 4146.39 & 219.272 & 0.022\\
  ASAS J112105-3845.3 & 0.18 & 3248.14 & 0.23 & 3283.66 & 0.24 & 3321.61 & 464.318 & 0.007\\
  ASAS J112117-3446.8 & 0.71 & 3962.67 & 0.7 & 3719.79 & 0.84 & 4013.62 & 269.414 & 0.02\\
  ASAS J112205-2446.7 & 1.41 & 5111.29 &  &  &  &  &  & \\
  ASAS J115942-7601.4 & 0.75 & 4021.5 & 0.79 & 3729.38 & 0.86 & 4131.34 & 398.782 & 0.02\\
  ASAS J120139-7859.3 & 1.81 & 4936.73 &  &  &  &  &  & \\
  ASAS J120204-7853.1 & 0.76 & 4036.83 & 0.79 & 3730.57 & 0.88 & 4162.21 & 398.782 & 0.011\\
  ASAS J121138-7110.6 & 1.7 & 4828.8 &  &  &  &  &  & \\
  ASAS J121531-3948.7 & 0.75 & 3996.59 & 0.74 & 3747.22 & 0.87 & 4076.6 & 244.451 & 0.021\\
  ASAS J122023-7407.7 & 0.71 & 3985.11 & 0.76 & 3709.42 & 0.83 & 4073.53 & 442.914 & 0.003\\
  ASAS J122034-7539.5 & 0.76 & 4304.98 & 0.78 & 4110.09 & 0.75 & 4436.18 & 103.142 & 0.034\\
  ASAS J122105-7116.9 & 0.69 & 3971.33 & 0.74 & 3700.36 & 0.82 & 4052.4 & 442.914 & 0.015\\
  ASAS J123921-7502.7 & 1.03 & 4276.23 & 0.98 & 3868.27 & 1.09 & 4514.39 & 283.183 & 0.014\\
  ASAS J125826-7028.8 & 1.19 & 4392.09 & 1.15 & 4041.77 & 1.23 & 4715.63 & 233.527 & 0.009\\
  ASAS J134913-7549.8 & 1.07 & 5532.3 & 0.98 & 5212.66 & 1.06 & 5745.5 & 27.07 & 0.084\\
  ASAS J153857-5742.5 & 1.18 & 5137.52 & 1.16 & 4707.41 & 1.14 & 5642.4 & 68.336 & 0.063\\
  ASAS J171726-6657.1 & 1.28 & 5678.07 & 1.27 & 5278.14 &  &  &  & \\
  ASAS J181411-3247.5 & 1.13 & 4912.49 & 1.11 & 4520.86 & 1.09 & 5366.58 & 86.862 & 0.028\\
  ASAS J181952-2916.5 & 1.22 & 5334.97 & 1.2 & 4899.31 &  &  &  & \\
  ASAS J184653-6210.6 & 0.7 & 3955.99 & 0.71 & 3755.56 & 0.78 & 3955.37 & 222.362 & 0.024\\
  ASAS J185306-5010.8 & 1.18 & 5144.63 & 1.16 & 4714.0 & 1.14 & 5651.12 & 68.336 & 0.014\\
  ASAS J200724-5147.5 & 0.59 & 3897.56 & 0.6 & 3684.73 & 0.62 & 3826.1 & 184.707 & 0.005\\
  ASAS J204510-3120.4 & 0.54 & 3806.89 & 0.58 & 3625.95 & 0.62 & 3721.1 & 314.173 & 0.015\\
  ASAS J205603-1710.9 & 0.92 & 4271.71 & 0.9 & 3991.7 & 0.91 & 4447.94 & 147.384 & 0.023\\
  ASAS J212050-5302.0 & 0.95 & 4933.09 & 0.95 & 4573.6 & 0.99 & 5233.2 & 49.717 & 0.069\\
  ASAS J214430-6058.6 & 0.62 & 3896.88 & 0.67 & 3758.79 & 0.65 & 3823.77 & 200.199 & 0.023\\
  ASAS J232749-8613.3 & 0.95 & 4953.28 & 0.95 & 4587.34 & 0.99 & 5252.98 & 49.717 & 0.014\\
  ASAS J233231-1215.9 & 0.65 & 3904.13 & 0.67 & 3709.93 & 0.73 & 3859.11 & 254.116 & 0.022\\
  ASAS J234154-3558.7 & 1.0 & 5317.63 & 0.96 & 5156.84 & 0.99 & 5696.24 & 38.991 & 0.046\\
  SWASP1 J002334.66+201428.6 & 0.47 & 3732.26 & 0.51 & 3563.56 & 0.55 & 3642.87 & 350.494 & 0.022\\
  SWASP1 J021055.38-460358.6 & 0.79 & 4561.22 & 0.76 & 4476.24 & 0.78 & 4857.86 & 62.811 & 0.018\\
  SWASP1 J022729.25+305824.6 & 0.78 & 4043.37 & 0.78 & 3822.18 & 0.83 & 4103.71 & 195.566 & 0.064\\
  SWASP1 J033120.80-303058.7 & 1.37 & 6297.53 & 1.31 & 6078.61 &  &  &  & \\
  SWASP1 J041422.57-381901.5 & 1.2 & 5870.07 & 1.06 & 5223.15 & 1.13 & 5981.71 & 21.575 & 0.078\\
  SWASP1 J042148.68-431732.5 & 1.29 & 6071.67 & 1.24 & 5882.23 &  &  &  & \\
  SWASP1 J043450.78-354721.2 & 0.82 & 4289.73 & 0.83 & 4071.42 & 0.82 & 4456.16 & 118.681 & 0.019\\
  SWASP1 J045153.54-464713.3 & 0.96 & 4985.07 & 0.96 & 4609.35 & 1.0 & 5283.93 & 31.85 & 0.089\\
  SWASP1 J050649.47-213503.7 & 0.45 & 3704.11 & 0.49 & 3541.96 & 0.52 & 3616.87 & 405.198 & 0.033\\
  SWASP1 J051829.04-300132.0 & 0.79 & 4195.07 & 0.8 & 3994.46 & 0.78 & 4318.64 & 146.045 & 0.012\\
  SWASP1 J052855.09-453458.3 & 0.74 & 4105.35 & 0.77 & 3930.63 & 0.75 & 4167.13 & 146.045 & 0.032\\
  SWASP1 J053504.11-341751.9 & 0.73 & 4368.41 & 0.71 & 4293.5 & 0.72 & 4568.21 & 70.363 & 0.016\\
  SWASP1 J054516.24-383649.1 & 0.82 & 4293.94 & 0.83 & 4081.86 & 0.82 & 4455.43 & 118.681 & 0.011\\
  SWASP1 J055021.43-291520.7 & 1.03 & 5349.2 & 1.0 & 4855.19 & 1.06 & 5597.84 & 24.489 & 0.142\\
  SWASP1 J064118.50-382036.1 & 0.79 & 4592.53 & 0.77 & 4506.23 & 0.79 & 4908.82 & 62.811 & 0.012\\
  SWASP1 J101828.70-315002.8 & 0.85 & 4101.65 & 0.83 & 3819.12 & 0.94 & 4253.52 & 219.272 & 0.002\\
  SWASP1 J114824.21-372849.2 & 0.51 & 3775.15 & 0.52 & 3575.22 & 0.62 & 3714.79 & 423.631 & 0.012\\
  SWASP1 J191144.66-260408.5 & 1.16 & 5037.76 & 1.14 & 4621.4 & 1.12 & 5518.4 & 68.336 & 0.084\\
  SWASP1 J224457.83-331500.6 & 0.39 & 3645.74 & 0.44 & 3499.12 & 0.47 & 3567.25 & 487.059 & 0.005\\
  SWASP1 J231152.05-450810.6 & 1.04 & 5472.7 & 1.01 & 5301.17 & 1.04 & 5873.77 & 32.162 & 0.16\\
\end{longtable}
\end{longtab}

In  Fig. \ref{dotemp}, we plot   $\Delta\Omega_{\rm phot}$  vs. the effective temperature $T_{\rm eff}$ inferred from the \citet{2000A&A...358..593S}  isochrones.
The model of \citet{2011AN....332..933K} predicts the relationship $\Delta\Omega = 0.071 (\frac{T_{\rm eff}}{5500})^2$  for ZAMS stars with $T_{\rm eff} \le 6000 \rm K$.
Our targets comprise PMS stars that are still contracting and stars recently settled on MS. We select MS stars by comparing the masses and ages of our targets with the values tabulated by \citet{2000A&A...358..593S} for the ZAMS stars. 
We fit our ZAMS data with a power-law and we find $\Delta\Omega_{\rm phot}=0.09(\frac{T_{\rm eff}}{5500})^{2.18\pm0.65}$. 

The use of  \citet{1998A&A...337..403B} or  \citet{2013ApJ...776...87S}  isochrones does not change significantly this result. The power-laws obtained with the two models are $\Delta\Omega_{\rm phot}=0.105(\frac{T_{\rm eff}}{5500})^{2.06\pm 0.7}$ and  
$\Delta\Omega_{\rm phot}=0.09(\frac{T_{\rm eff}}{5500})^{2.6\pm0.6}$, respectively.
Though the data are quite scattered, the exponents of the theoretical  and fitted power laws are in good agreement with each other. However, the two curves exhibit a small offset of about $0.01~\rm rad~d^{-1}$ and our measurements are, on average, higher than the values predicted by \citet{2011AN....332..933K}. We remind that our analysis tends to underestimate SDR, hence the shift between theoretical and real values could be more pronounced than that shown in the picture. 
In Fig. \ref{dotemp}, we also plot   the results of \citet{2013A&A...560A...4R} and those from \citet{2005MNRAS.357L...1B} for comparison.
We plot the median $\Delta\Omega_{\rm phot}$ values found by \citet{2013A&A...560A...4R} in different $T_{\rm eff}$ bins. These median values fall inside the $95\%$ confidence region of our fitted power law. Hence the two works can be considered in agreement.
Note however that the trend  found  by \citet{2013A&A...560A...4R} is  more  flattened than the power law found here and  that predicted by \citet{2011AN....332..933K}. This discrepancy could be due to the fact that \citet{2013A&A...560A...4R} mix stars with different ages and with a wider range of rotation periods.

\citet{2005MNRAS.357L...1B} analyze a sample of ten young late-type stars and find the relationship $\Delta\Omega\propto T_{\rm eff}^{8.92}$. The targets investigated by \citet{2005MNRAS.357L...1B} comprises PMS stars and stars recently settled in the ZAMS.
Their power law  is much more steep than that predicted by \citet{2011AN....332..933K} and in disagreement with the trend of our PMS and ZAMS data.  Hence our targets do not confirm \citet{2005MNRAS.357L...1B}  finding.

\subsubsection{Correlation between SDR and  global convective turnover time-scale}
The stars analyzed in the present work span the age range $4-95~ \rm Myr$. During this time interval, the stellar structure evolves from a fully convective structure to a radiative core plus a convective envelope. 
Hence,  in our sample, stars with the same effective temperature can have very different structures depending on their age and mass.  
In this age range, the convective turn-over time scale $\tau_C$ could be a more convenient parameter for investigating the relationship  between the SDR and the stellar structure.
Indeed $\tau_C$ is proportional to the depth of the convective zone and is more representative of the stellar structure than the effective temperature.     

\begin{figure*}
\begin{center}
\includegraphics[angle=0]{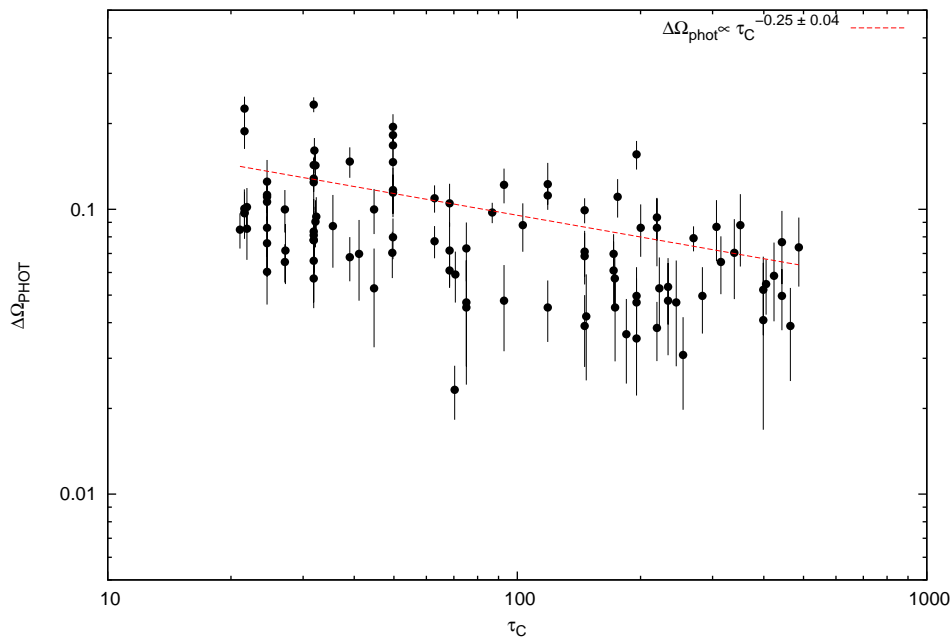}
\caption{ $\Delta\Omega_{\rm phot}$ vs. the convective turn-over time-scale $\tau_C$. }
\label{taudo}
\end{center}
\end{figure*}

In Fig. \ref{taudo} we plot $\Delta\Omega_{\rm phot}$ vs. $\tau_C$. 
$\Delta\Omega_{\rm phot}$ increases toward  shorter $\tau_C$ values.  Hence $\Delta\Omega_{\rm phot}$ increases as the depth of the convective envelope decreases. This  result is in agreement with the theoretical models  of  \citet{1999A&A...344..911K}, \citet{2011AN....332..933K} and  \citet{2011A&A...530A..48K}.
We fit a power law to our data and we find $\Delta\Omega_{\rm phot}\propto\tau_C^{-0.25\pm 0.04}$ (red line in the picture).
Though the data exhibit a clear trend, they are broadly spread around the fitted power law. 
This spread is partly due to the limitations of our measurement method and partly due to the fact that the plot mixes stars with different rotation periods.

 \subsubsection{Correlation between SDR and the rotation period}
\begin{figure*}
\begin{center}
\includegraphics[angle=0]{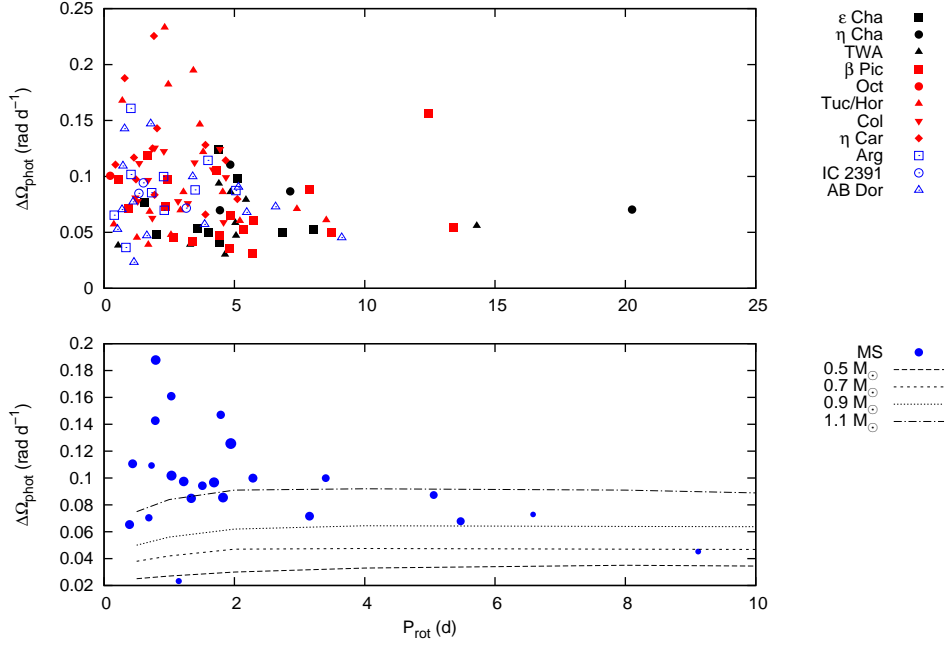}
\caption{Top panel:  $\Delta\Omega_{\rm phot}$ vs. the stellar rotation period $P_{\rm rot}$. The highest $\Delta\Omega_{\rm phot}$ values have been measured in stars with $P_{\rm rot}\le5 \rm d$.  Bottom panel: the same plot restricted to the MS stars.  The symbol sizes are proportional to the stellar masses that range between $0.6$ and $1.35~\rm M_{\sun}$.  The black lines are the rotational tracks derived by \protect\citet{2011AN....332..933K} for stars of,$0.5$, $0.7$, $0.9$ and $1.1 ~\rm {M}_{\sun}$ as labelled, respectively.  }
\label{rotdo}
\end{center}
\end{figure*}

 In the top panel of Fig. \ref{rotdo} we plot $\Delta\Omega_{\rm phot}$ vs. the rotation period $P_{\rm rot}$.
In this case, the data do not follow a well defined trend and cannot be reproduced by a power law.
However, SDR seems somehow related to the stellar rotation period. In fact, the highest values of  $\Delta\Omega_{\rm phot}$ correspond to $P_{\rm rot}$ between $0.7$ and $5~\rm d$.
\citet{2011AN....332..933K} study in detail how SDR depends on rotation period in ZAMS stars. They compute differential rotation for  stars  with different masses and periods and derive a set of "rotational tracks". Each track shows how $\Delta\Omega$ depends on rotation period for a fixed stellar mass. In  the bottom panel of Fig. \ref{rotdo}, we plot the rotational tracks derived by \citet{2011AN....332..933K} for  stars with $0.5, 0.7, 0.9$ and $1.1~\rm  M_{\sun}$. According to these tracks, SDR is almost independent of the stellar rotation period and is more influenced by the stellar mass.
We over-plot our results  on the tracks for comparison. We report only stars that have reached the ZAMS  because the tracks have been computed for ZAMS stars. The masses of these targets cover the range $0.6-1.35~\rm M_{\sun}$. 
The circles size is proportional to the stellar mass.
Our results show a trend that partly resembles the rotational tracks. Indeed, the less massive stars exhibit on average  a lower SDR.
However, for stars with $P_{\rm rot}$ between 0.7 and 2 d, there is a strong discrepancy with the model predictions. This suggests that the stellar mass cannot be the main parameter on which SDR depends as claimed by \citet{2011AN....332..933K} and that stellar rotation period plays  a key role too.

\subsubsection{Correlation between SDR and  stellar age}

\begin{figure*}
\begin{center}
\includegraphics[angle=0]{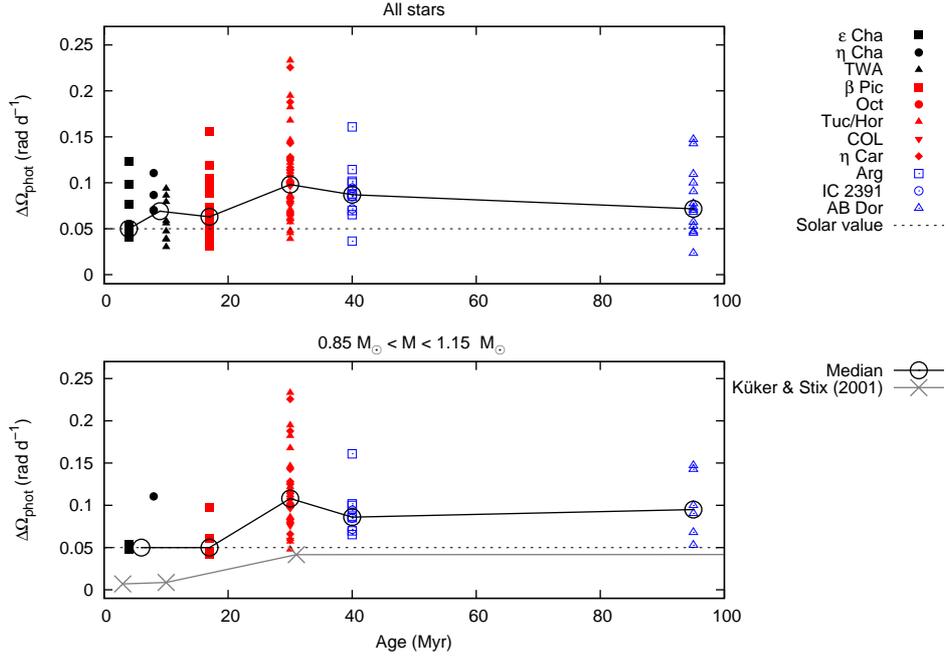}
\caption{Top panel:
 $\Delta\Omega_{\rm phot}$ vs. the stellar age.The empty circles indicate the median $\Delta\Omega_{\rm phot}$ values measured at different ages (note that the members of $\eta~Cha$ have been grouped with the members of $\rm TWA$).
The black continuous line connects the median $\Delta\Omega_{\rm phot}$ values and the dotted line marks the present solar shear for comparison.  Bottom panel: the same plot restricted to stars with mass between $0.9$ and $1.1~\rm M_{\sun}$.  The grey crosses mark the rotational shear for the Sun at different ages as predicted by  \protect\citet{2001A&A...366..668K} }
\label{dotime}
\end{center}
\end{figure*}

The stellar associations studied in the present work span an age range between 4 and 95 Myr.
Hence our sample includes stars that are in the first phases of the PMS stage and stars that are approaching or recently settled on the MS.
Our results are therefore  useful to investigate how SDR evolves in time during the PMS stage.
In the top panel of Fig. \ref{dotime}, we display the measured values of $\Delta\Omega_{\rm phot}$ vs. the stellar age. 
The median $\Delta\Omega_{\rm phot}$ increases between 4 and 30 Myr  and then decreases and approaches the current solar value at 95 Myr. 
The scatter of the data is partly due  to the intrinsic limitations related to our methodology  and partly to the fact that the picture mixes stars with different masses and rotation periods.
In the bottom panel of Fig. \ref{dotime} we report only stars with masses between 0.85 and 1.15 $\rm M_{\sun}$ and we  investigate how $\Delta\Omega_{\rm phot}$ evolves in a young Sun. The median $\Delta\Omega_{\rm phot}$ is constant in the first $17~ \rm Myr$, then it  significantly  increases and,  at $30~\rm  Myr$, its value is about twice the initial value. Between 30 and 95 Myr $\Delta\Omega_{\rm phot}$ remains almost constant.
\citet{2001A&A...366..668K} model  SDR evolution in the Sun. They develop four theoretical models for the Sun at $3$, $10$ and $31~\rm Myr$ and for the present Sun. 
In their models $\Delta\Omega$ is inversely correlated to the depth of convective zone. This implies that $\Delta\Omega$ increases between 3 and 31 Myr. At 3 Myr, the Sun has a fully convective structure and  a small SDR. As the Sun evolves in time, a radiative core grows, the depth of the convective zone decreases and $\Delta\Omega$ increases. \citet{2001A&A...366..668K} do not find a significant difference between the 31 Myr  and the present Sun models. Hence, according to their work, the Sun should have reached its current $\Delta\Omega$ value at about 30 Myr.
Also in this case, our results agree only qualitatively with the model prediction. Indeed the median $\Delta\Omega_{\rm phot}$  values increase between 4 and 30 Myr,  but they are systematically higher than  those predicted by \citet{2001A&A...366..668K}. 
Note that, though these young stars exhibit  a SDR greater than the solar one, they can be regarded as solid body rotators because they rotate faster than the present Sun and have a low relative shear $\alpha=\frac{\Delta\Omega}{\Omega_{\rm eq}}$

\section{Discussion}
In the last section we showed that our results are qualitatively in agreement with the models developed by \citet{1999A&A...344..911K} and \citet{2011AN....332..933K}.
However our $\Delta\Omega_{\rm phot}$ measurements are often  higher than the values predicted by the same models (see Fig. \ref{dotemp} and \ref{rotdo}), especially in  stars with periods between 0.7 and 2 d.
Such a disagreement between $\Delta\Omega$ values and models predictions has been already noticed and discussed by \citet{2006MNRAS.370..468M}. 
These authors measured an absolute shear $\Delta\Omega=0.4~ \rm rad~d^{-1}$ for the $G0$ star $\rm HD ~171488$. 
This value is  significantly higher than those measured for the  G dwarfs $\rm R58$ and $\rm LQ~ Lup$, that are $\Delta\Omega=~0.138 \rm rad~d ^{-1}$ and $\Delta\Omega=0.12 ~\rm rad~d^{-1}$, respectively  \citep{2005ESASP.560..799M,2000MNRAS.316..699D}.
According to the models, the three stars should have about the same rotational shear because the stellar temperature is the main parameter on which $\Delta\Omega$ depends.
 \citet{2011A&A...530A..48K} were not able to find an explanation about this discrepancy and questioned the reliability of the $\Delta\Omega$ measurement for $\rm HD~ 171488$. However \citet{2008MNRAS.390..635J} made an independent measurement  and found $\Delta\Omega = ~0.5 \rm rad~d^{-1}$
The main difference between $\rm HD~ 171488$ and the other two stars lies in the rotation period. Indeed, $\rm HD~ 171488$ has a rotation period of $1.31~\rm d$, whereas the other two stars have a rotation period of 0.5 and 0.3 days respectively. 

As a matter of fact, the rotational tracks computed by \citet{2011AN....332..933K} predict that SDR amplitude slightly increases with the period in the range 0.3-2 d (see bottom panel of Fig. \ref{rotdo}) but this increment is too small to explain the SDR amplitude of HD 171488.   
The  high differential rotation found in  $\rm HD~ 171488$ and in our targets, with periods between 0.7 and 2 d, suggest that the dependence on the rotation period could be more pronounced than models prediction.

A stronger dependence on the stellar rotation rate could also explain the discrepancy between the power law found by \citet{2005MNRAS.357L...1B} and that found in the present work (see Fig \ref{dotemp}).
Indeed, the high steepness of the power law found by \citet{2005MNRAS.357L...1B} is due to the M1V stars HK Aqr and EY Dra for which the authors measure very low SDR values i.e. 0.005 and $0.0003~\rm rad~d^{-1}$, respectively.
The M1V stars of our sample have in comparison higher $\Delta\Omega_{\rm phot}$ values.
However, we  point out that HK Aqr and EY Dra  rotate in about 0.5 days whereas the M1V stars investigated here have rotation periods between 1.5 and 5.5 days. 
Thus the difference between HK Aqr, EY Dra and the M1V stars investigated here could be due to the different rotation periods.

\begin{figure*}
\begin{center}
\includegraphics[angle=0]{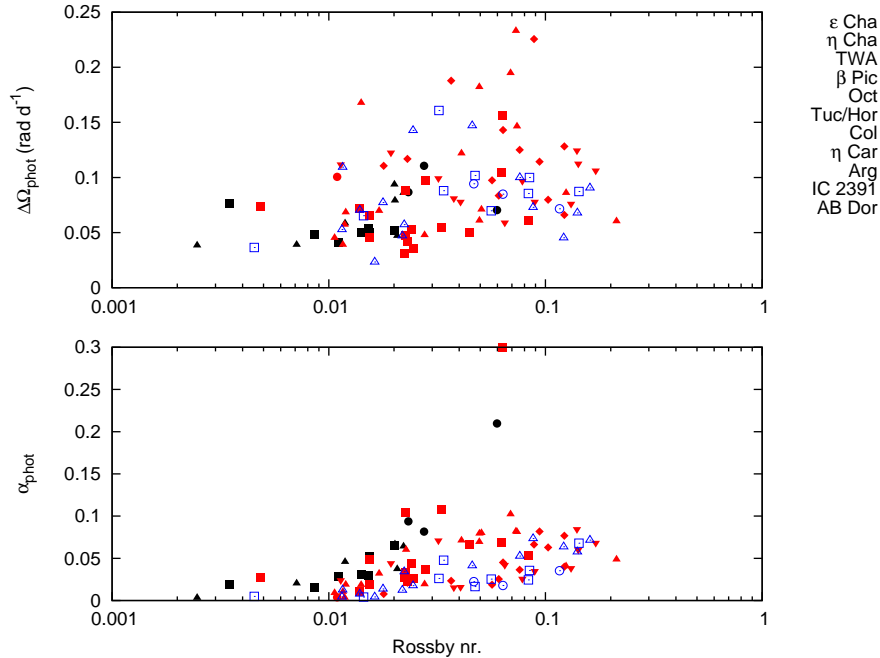}
\caption{Top panel: the absolute shear $\Delta\Omega$ vs. the Rossby number Ro. Bottom panel: the relative shear $\alpha$ vs. Ro. $\Delta\Omega$ and $\alpha$ increase toward higher Ro values. This trend is an agreement with the model developed by \protect\citet{2014MNRAS.438L..76G}.}
\label{rossbydo}
\end{center}
\end{figure*}

The models developed by \citet{2011AN....332..933K} are based on the so called $\Lambda$ effect.
In these models the main driver of the differential rotation is the non-diffusive term of the Reynolds stress induced by the interaction between the Coriolis force and the convective motions.
As noticed by \citet{2014MNRAS.438L..76G}, these models depend on free parameters like the  turbulent viscosity coefficients.
A different choice of these parameters could maybe lead to a stronger dependence on stellar rotation rate.
\citet{2014MNRAS.438L..76G} developed a model based on a 3D hydrodynamical code where the Reynolds stresses do not need to be parametrized.
In this model, the main parameter on which $\Delta\Omega$ depends is the Rossby number $Ro=\frac{P_{\rm rot}}{\tau_{\rm c}}$ where $P_{\rm rot}$ is the stellar rotation period and $\tau_{\rm c}$ is the convective turnover timescale.
If $Ro< 1$, that is the case of our targets,  the Coriolis force dominates the buoyancy and the star  tends to rotate as a solid-body.  As Ro increases the Coriolis force becomes less important and the value of differential rotation increases. For $Ro >1$ the buoyancy dominates on Coriolis force and the SDR becomes anti solar.
In Fig. \ref{rossbydo} we plot $\Delta\Omega_{\rm phot}$ vs. the Rossby Number  computed in Sec. 3. The plot is quite scattered, but the trend of $\Delta\Omega_{\rm phot}$ seems to confirm the prediction of \citet{2014MNRAS.438L..76G}.
In the bottom panel of the Fig. \ref{rossbydo}, we report the relative shear $\alpha_{\rm phot}$ vs. the Rossby number.
The trend and the range of $\alpha_{\rm phot}$ values are very similar to that predicted by \citet{2014MNRAS.438L..76G} and shown in the Fig. 2 of their work.

\section{Conclusions}
\label{sec:summary}
We investigate the correlation between the amplitude of SDR and global stellar parameters in members of young loose stellar associations.
We measure the quantities  $\Delta\Omega_{\rm phot}$ and $\alpha_{\rm phot}$ for 111 stars  by processing long-term photometric time-series. These quantities are lower limits to the absolute and to the relative surface rotational shear, respectively.
Our analysis leads to the following results:

\begin{itemize}

\item{ $\Delta\Omega_{\rm phot}$ increases with the effective temperature $T_{\rm eff}$ following the power law $\Delta\Omega_{\rm phot}\propto T^{2.18\pm 0.65}_{\rm eff}$ in MS stars. This power law is very close to that predicted by \citet{2011AN....332..933K} (i.e. $\Delta\Omega\propto T^2_{\rm eff}$);  }

\item{the PMS stars of our sample show a trend very similar to that exhibited by MS stars. Thus, the power law $\Delta\Omega \propto T^{8.6}_{\rm eff}$ found   by \citet{2005MNRAS.357L...1B} for  PMS and ZAMS stars is  not confirmed by our measurements;}

\item{ $\Delta\Omega_{\rm phot}$ increases with decreasing convective turnover time-scale $\tau_C$  according to the power law $\Delta\Omega_{\rm phot}\propto\tau_C^{-0.25\pm 0.4}$;}

\item{our  $\Delta\Omega_{\rm phot}$ measurements are  systematically higher than the values predicted by \citet{2011AN....332..933K}. This discrepancy is particularly large in  stars with a rotation period between $0.7$ and $2$ d and suggests that the dependence on the rotation period could be stronger than the model prediction;}

\item{we investigate the time evolution of $\Delta\Omega_{\rm phot}$ for a $1 ~ \rm M_{\sun}$ stars and find that $\Delta\Omega_{\rm phot}$ increases with the stellar age in the first $30~ \rm Myr$. This is consistent with the theoretical models that predict a low degree of differential rotation for fully convective stars;}

\item{$\Delta\Omega_{\rm phot}$  and $\alpha_{\rm phot}$ increase with the Rossby number $Ro$ in agreement with the theoretical model developed by \citet{2014MNRAS.438L..76G}. }

\end{itemize} 

\begin{acknowledgements} 
The authors are grateful to Rainer Arlt and to the anonymous referee for helpful comments and suggestions. 
 \end{acknowledgements}

\label{lastpage}

\bibliographystyle{aa}
\bibliography{SDRref}
\end{document}